\documentclass[11pt]{article}

\usepackage{times,amsmath,amssymb}
\usepackage{graphicx} % for pdf figures
\usepackage{graphics} % for ps  figures
\usepackage{epsfig}
\usepackage{color}
\usepackage{mymacros}

\numberwithin{equation}{section}

\newtheorem{lemma}{Lemma}[section]

\usepackage[ruled]{algorithm2e}

\begin{document}
%\title{Cone Schedules for Processing Systems in Modulated Environments}

%\author{KEVIN ROSS
%\affil{University of California Santa Cruz}
%NICK BAMBOS
%\affil{Stanford University}
%GEORGE MICHAILIDIS
%\affil{The University of Michigan}
%}

\begin{center}
{\LARGE Cone Schedules for Processing Systems in Fluctuating Environments}
\vskip 0.5cm

KEVIN ROSS
\footnote{School of Engineering, University of California Santa Cruz; kross@soe.ucsc.edu; }\\
NICHOLAS BAMBOS
\footnote{Electrical Engineering and Management Science \& Engineering, Stanford University; bambos@stanford.edu}\\
GEORGE MICHAILIDIS
\footnote{Statistics and Electrical Engineering \& Computer Science, The University of Michigan; gmichail@umich.edu}

\end{center}
%
%~\vskip 0.5cm
%\begin{center}
%{\large\bf CONE SCHEDULES FOR PROCESSING SYSTEMS \\
%\vskip 0.3cm
%IN MODULATED ENVIRONMENTS\\}
%\vskip 0.5cm
%{\large Kevin Ross\footnote{kross@soe.ucsc.edu.edu; Baskin School of
%Engineering, University of California, Santa Cruz, CA 95064.},
%Nicholas Bambos\footnote{bambos@stanford.edu; Department of
%Management Science \& Engineering, and Department of Electrical
%Engineering, Stanford University, Stanford, CA 94305.} and George Michailidis\footnote{gmichail@umich.edu; Department of Statistics and Electrical Engineering \& Computer Science,
%The University of Michigan, Ann Arbor, MI 48109}
% }
%\date{}

%\vskip 0.5cm

\begin{abstract}

We consider a generalized processing system having several queues, where the available service rate combinations are fluctuating over time due to reliability and availability variations. The objective is to allocate the available resources, and corresponding service rates, in response to both workload and service capacity considerations, in order to maintain the long term stability of the system. The service configurations are completely arbitrary, including negative service rates which represent forwarding and service-induced cross traffic. We employ a trace-based trajectory asymptotic technique, which requires minimal assumptions about the arrival dynamics of the system.

We prove that {\em cone schedules}, which leverage the geometry of the queueing dynamics,  maximize the system throughput for a broad class of processing systems, even under adversarial arrival processes.
We
study the impact of fluctuating service availability, where resources
are available only some of the time, and the schedule must dynamically
respond to the changing available service rates, establishing both the capacity of such systems and the class of schedules which will stabilize the system at full capacity.  The rich geometry of the system dynamics leads to important insights for stability, performance and scalability, and substantially generalizes previous findings.

The processing system studied here models a broad variety of computer, communication and service networks, including varying channel conditions and cross-traffic in wireless networking, and call centers with fluctuating capacity. The findings have implications for bandwidth and processor allocation in communication networks and workforce scheduling in congested call centers. By establishing a broad class of stabilizing schedules under general conditions, we find that a scheduler can select the schedule from within this class that best meets their load balancing and scalability requirements.

%\footnote{Partial results have appeared in
%\cite{RoB:02a,RoB:02b,RoB:04a,RoB:05b,RoB:06a,RoB:09}}.

\end{abstract}

%\category{}{}{}

%\terms{}

%\keywords{random environment, stability, adversarial queueing theory, dynamic scheduling, throughput
%maximization.}
%\maketitle


\section{Introduction }
\label{secIntro}

We consider a processing system comprised of $Q$  infinite capacity queues, indexed by $q\in\mathcal{Q}=\{1,2,...,Q\}$, operating in a time-varying environment which fluctuates amongst environment states $e \in \cE = \{1,2,...,E\}$. In each environment state, only a subset of the service configurations are available. The process scheduler selects a service configuration vector $S=(S_1,\cdots,S_Q)$ from the environment-dependent available set $\cS^e$. Upon selection, if $S_q>0$ then queue $q$ is emptied at rate $S_q$, and if $S_q<0$ then the queue is filled at the corresponding rate. The available service configurations can be completely arbitrary, including vectors with any combination of positive and negative components.

A key question addressed in this study is which of the available service configurations should be
selected, given the system workload and environment state histories, so as to maximize its
throughput. We introduce a family of resource allocation policies - called Cone Schedules -
which are shown to stabilize the system under the maximal possible traffic load, even if that load is designed by an adversary to destabilize the system whenever possible.

This canonical processing model captures several applications in 
computing and communication systems, including wireless networks, packet switches and call centers. The main characteristic of these applications is that the service rates across multiple queues are coupled through operational constraints,
giving rise to the available service configurations.   Service rate availability (corresponding to the environment states) is affected by staff scheduling in call centers, congestion dynamics in wireless networks and scheduled or unscheduled outages due to maintenance or reliability issues in other processing systems.

\subsection{Related Work}
The trace-based stability analysis technique employed in this paper relates to the study of adversarial queueing networks exemplified in \cite{And:01} and \cite{Bor:01}. This approach avoids imposing a probabilistic framework on the arrival traffic, and instead analyzes the performance of a queueing network under and adversarial arrival traffic trace, designed to stress the system as much as possible.  They describe a queueing network as universally stable when they can show that the total workload of the system is bounded under any deterministic or stochastic adversary's arrival trace. This work is really finding the {\em worst-case} behavior of a network by considering the network to be a {\em game} between the schedule (protocol) and the worst possible arrival trace (adversary). They limit the absolute arrival volume within a finite interval, but do not require it to follow any stationary distribution or apply any further restrictions.
This concept builds upon earlier work called {\em leaky-bucket} analysis in \cite{Cruz:91a} and \cite{Cruz:91b}.  

Adversarial models have been used to in packet networks before, such as \cite{Bor:01} which considers a fixed-path packet network. Some more general queueing systems, including multiclass queueing networks are studied in \cite{Tsa:00}, with generalized service times and heterogeneous customers.  Adversarial methods have  also been employed to study multi-hop network stability in \cite{Kushner:06}. In \cite{Ans:02}, adversarial models are used to analyze load-balancing algorithms in a distributed setting based using a token-based system on a network with limited deviations from the average load. While none of these study the same network scheduling setting of this paper (to our knowledge they have only considered fixed-path networks under time-invariant service environments), each example presents a persuasive argument for the value of network stability analysis in the absence of a well-defined probabilistic framework.

A special example of the system described in this paper is a single crossbar
packet/cell switch with virtual output queues, used in high speed IP
networks.  The switch paradigm is the focus of \cite{RoB:09}, and
provides a helpful context to develop the cone algorithms. In this
switch, cells arriving to each input port get buffered in separate
virtual queues, based on the output port they are destined to. The
switching fabric can be set to a different connectivity mode  in
each time slot, matching each input port with a corresponding output
port for cell transfer.  In this context, {\em Maximum Weight
Matching} (MWM) has been shown in
\cite{McKw:99} to maximize the throughput of input queued switches,
employing Lyapunov methods for stability analysis, as also in
constrained queueing systems studied in
\cite{TasE:92,Tass:95,TassB:00,HM:11}.  In
our more general service model, MWM corresponds to maximizing $\<
S,X\>=\sum_q S_q X_q$, where the weight $X_q$ is the cell
workload of queue $q$ or a related congestion measure, and the $S$ vectors represent the crossbar configurations.

More general results on the stability of MWM algorithms, using fluid
scaling methods, were later obtained in \cite{DaiP:00},  and on a generalized switch model in
\cite{Stol:04}.  Stability in networks of switches was studied in \cite{Marsan:05} and 
\cite{Leonardi:05}.   \cite{DaiLin:05} and \cite{DaiLin:08} considered maximum
pressure policies by modeling fluid flows for types of processing
networks. Their work can be seen as a generalization of the  policies which maximize $\sum_q S_q X_q$ where some of the service rates are negative because the available configurations involve forwarding workload from one queue to another downstream queue. 
\cite{Neely:05}  studied broader optimal controls for generalized
(wireless) network models that involve joint scheduling, routing and
power allocation. All of these have significantly advanced the theory of the stability of scheduling rules which allocate service to queues based on a weighted-matching approach, and utilize a probabilistic framework to apply fluid limit or heavy-traffic analysis.

Instead of using fluid scaling methods (primarily analytic,
involving passage to a limit regime) to establish the results, we
opt to use an alternative direct and primarily geometric approach
in this work, which seems to have broader applicability to other
queueing systems and reveals useful geometric insight regarding
their dynamics.
The trace-based asymptotic analysis employed here was introduced in
\cite{ArmB:03}, where  the maximum weight matching algorithms were studied and it was shown that maintain maximal throughput is guaranteed under very general arrival process assumptions.  The method was also employed in  \cite{BM:04} where randomly fluctuating service levels were studied.  In that case the service rate assignments are made without full knowledge of service availability, as opposed to the processing systems studied here where service allocation decisions are made in response to availability. Like the adversarial queueing models, there is no probabilistic framework required, but unlike the traditional adversarial models, there is also no short-term restriction on arrival bursts in finite time, but just a long-term traffic load restriction. This leads to more general stability results, but eliminates the possibility of tighter bounds on other performance metrics. For example under such general assumptions there can be no guaranteed finite bound on the total workload, or even the expected workload in the system.

\subsection{Results Overview}
                    
We classify the stability region for these processing systems with fluctuating service availability.  We find that rate stability  for these general processing systems can be guaranteed by the class of {\em cone schedules}, for any arbitrary arrival process that can possibly be stabilized.
 Cone schedules use the available 
service vector with maximal projection $\< S,\bB
X\>=\sum_p\sum_q S_pB_{pq}X_q$ on the {\em projected} workload vector
$\bB X$, for every matrix that is {\em positive-definite}, has {\em
negative or zero off-diagonal} elements, and is {\em symmetric}. This substantially generalizes a similar result in \cite{RoB:09} , where the same class of algorithms was shown to maximize throughput for the special case of packet switches. 

In classifying the stability region, we show how the combination of service vectors in each environment impacts the overall capacity of the system, beyond the long term availability of each service vector. The geometric framework for stability aids the intuition and analysis significantly. Because of environment fluctuations, one may expect that a scheduling rule needs to account for future and past states. However we find that cone schedules, which respond only to the current workload, are able to guarantee stability for any arrival rate within the stability region.

The service rates in this paper are allowed to be completely arbitrary, in contrast to previous results using the trace-based analysis which  only applied to positive-service switches. This captures cross-traffic and forwarding between queues, because the selected service vector may induce additional workload to the system, in addition to the external arrival process. 
Further, in this work time is continuous, and arbitrarily large arrival bursts can be handled at arbitrarily small time intervals. This is more general than previous models where arrivals and decisions were restricted to timeslots.

From an architectural point of view, the geometric
approach to the scheduling problem provides key practical design
leads. Specifically, the conic representation (Section
\ref{secSchedules}) of cone schedules leads to scalable
implementations in switching systems.  Further, varying the elements of matrix
$\bB$, we can generate a very rich family of cone schedules that implement a soft {\em coupled
priority scheme} (and coupled load balancing) across the various
queues, managing delay tradeoffs between them. The schedules are also robust to any sublinear perturbation such as delayed or flawed state information.

The remainder of the paper proceeds as follows.  In Section \ref{secModel}, we introduce the model and system dynamics.  Section \ref{secStability} describes the throughput capacity or stability region of these networks, and in section \ref{secSchedules} we introduce the family of Cone Schedules and their geometry. Stability and performance implications are discussed in sections  \ref{secProof} and \ref{secPerf} respectively.  We conclude in Section \ref{secConclusions}.

\section{The Processing Structure}
\label{secModel}

Let $\int_{0}^{t} A_q(z)dz$ be the total workload that arrives to queue $q$ in the time interval $(0,t]$; that is, $A_q(t)\geq 0$ is the  instantaneous workload arrival rate at time $t\geq 0$. The traffic trace $\bA_q=\{A_q(t), t\geq 0\}$ is a (deterministic) function, which may have {\em discontinuities} and even {\em $\delta$-jumps} for each $q\in\cQ$. The overall (vector) instantaneous traffic rate is $A(t)=(A_1(t),A_2(t), ..., A_q(t), ..., A_Q(t))$ at time $t>0$ and the {\em traffic trace} is $\bA=\{A(t), t\geq 0\}$. We assume that the (long-term) {\em traffic load} of the trace\footnote{Throughout this study we employ the notation
$\ints_+=\{1,2,3,\cdots\}, ~\ints_{0+}=\{0,1,2,\cdots\},~ \reals=(-\infty,\infty),~\reals_+
=(0,\infty),~\reals_{0+}=[0,\infty)$}   $\bA$,
\eq
\label{eqratestab}
\lim_{t \go \infty} \frac{\int_{0}^{t}A(z)dz}{t}=\rho (\bA) \in \reals_{0+}^Q,
\en
is well-defined on the traffic trace $\bA$. Correspondingly, we define the set of traffic traces of load $\rho\in\reals_{0+}^Q$,
\eq
\label{trtrst}
\mathfrak{A}(\rho)=\left\{\bA=\{A(t),t\geq 0\} :
\lim_{t \go \infty} \frac{\int_{0}^{t}A(z)dz}{t}=\rho  \right\} ,
\en
restricting our attention in this paper to traffic traces of well-defined load. A variety of natural arrival processes are included. For example, $A_q(t)=\sum_{j=1}^\infty  \sigma_q^j\ \ind{t=t_q^j}$ models jobs of service requirement $\sigma_q^j$ arriving at times $t_q^j > 0$ to queue $q$. In this case, $A_q(t)$ is zero between consecutive $\delta$-jumps. In general, there could be {\em positive} instantaneous workload arrival rate between consecutive $\delta$-jumps, which would represent a continuous inflow of work.

No further restrictions are placed on the arriving traffic trace. It may be generated by an underlying stochastic process, or even an adversary specifically designed to destabilize the system whenever possible.

The arriving workload is queued up in the queues $q\in\cQ$, which are assumed to be of infinite capacity. Let $X_q(t)$ be the workload (total workload or service requirement) in queue $q$ at time $t\geq 0$ and
\eq
\nonumber
X(t)=(X_1(t),X_2(t),...,X_q(t),...,X_Q(t))\in \reals_{0+}^Q,
\en
the overall (vector) workload.

The processing system operates in a {\em fluctuating environment}, which can be in one of $E$ distinct states at any point in time, indexed by $e\in\cE=\{1,2,\cdots,E\}$. Let $e(t)\in \cE$ be the environment state at time $t$ and $\bE=\{e(t),~t\in\reals\}$ the overall environment trace over time. It is assumed that the proportion of time the environment trace $\bE$ spends in each state $e\in\cE$ is well-defined, that is,
\eq
\nonumber
\label{rate-env}
\lim_{t\go\infty} \frac{\int_0^t \ind{e(z)=z}dz}{t}=\pi^e(\bE)
\en
with $\sum_{e\in\cE}\pi^e(\bE) = 1, \pi^e (\bE)>0, e\in\cE$.  Correspondingly, we define the set of environment traces $\bE$ with time proportions $\pi^e, e\in\cE$ as
\eq
\label{envtrst}
\mathfrak{E}(\pi^e, e\in\cE )=\left\{ \bE=\{e(t),t\geq 0\}:
\lim_{t\go\infty}
\frac{\int_0^t \ind{e(z)=z}dz}{t}=\pi^e, e\in\cE \right\} ,
\en
and restrict our attention in this paper to environment traces that have well-defined time proportions. Finally, $E=1$ naturally corresponds to the degenerate case of a constant (non-fluctuating) environment.

When the environment is in state $e\in\cE$, a (nonempty) set of {\em service vectors} $\cS^e$ becomes available to the system manager, who can select a service vector $S\in\cS^e$ at any point in time to operate the system. Each $S\in\cS^e$ is a $Q$-dimensional vector
\eq
\nonumber
S=(S_1,S_2,...,S_q,...,S_Q)\in\reals^Q ,
\en
where $S_q\in\reals$ is the drain (or fill, see below) rate of queue $q$ when the service vector $S$ is used.  For example, in a simple system with two queues ($Q=2$), a service vector $S^1=(1.35,2.17)$ would serve (drain) queue 1 at rate 1.35 and queue 2 at rate 2.17 (work units per time unit). This is the standard way of viewing service vectors.

In this general model, however, we also allow for {\em negative} `service' rates, actually corresponding to traffic workload `feed' rates, as explained below.  In the previous simple example of two queues, a service vector $S^2=(1.2,-0.8)$ would serve (drain) the first queue at rate 1.2, but feed workload to the second queue at rate 0.8, filling it up.

The motivation to allow for negative components $S_q < 0$ in the service vectors $S\in\cS^e$ comes from the need to model {\em environmental (background) cross-traffic} sharing the queue buffers with the primary (foreground) traffic $\{A(t),t\geq 0\}$. This cross-traffic depends explicitly on the  service vector $S\in\cS^e$ used, and implicitly on the environment state $e\in\cE$ through the set $\cS^e$ where the service vector $S$ should be chosen from. When service vector $S$ is used with $S_q<0$ for some queue $q\in\cQ$, this corresponds to cross-traffic workload fed into queue $q$ at constant rate $-S_q > 0$, {\em in addition} to the primary traffic workload $\{A_q(t), t\geq 0\}$. It is easy to see that $- S_q >0$ can be interpreted as the `net' cross-traffic through the queue; that is, workload could be fed into queue $q$ at rate $r_1>0$  and removed (served) at rate $r_2>0$, with the net cross-traffic load fed into the queue being $-S_q=r_1-r_2 $.

One special case related to this model is a feed-forward network. A service vector representing the transfer of workload from one upstream queue $q^u$ to another downstream queue $q^d$ would be represented with $S_{q^u}=-S_{q^d}$ and all other $S_q=0$. The model here could handle the aggregate of many transfers, as well as gain and loss in the system at any queue. The concept of cross-traffic considered here is more general, requiring no restrictions on the physical structure of the network. Feed-forward networks require some additional assumptions and are not the primary focus of this paper, but have been studied extensively elsewhere, such as \cite{DaiLin:05}.

Note that the above environmental cross-traffic is far less `innocuous' than simply allowing the primary traffic to be modulated\footnote
{Actually, the environment could also modulate the primary traffic trace $\{A(t), t\geq 0\}$ in the following sense. There is a collection of traffic traces $\{A^e(t),t\geq 0\}$ one for each environment state $e\in\cE$. When the environment is in state $e$, the traffic driven into the system is selected from $\{A^e(t),t\geq 0\}$. Therefore, the overall traffic trace is simply $\{A(t)=\sum_{e\in\cE} A^e(t)\ind{e(t)=e}, t\geq 0\}$. Hence, this basically reverts to the standard model (as long as the limit $\lim_{t\go\infty} \int_0^t A(z)dz / t$ exists) and this is why we do not treat this case explicitly.}
by the environment state. Indeed, the cross-traffic depends on the choice of service vector $S$, hence, the scheduling decisions actively influence it. The environment plays only a secondary role by defining $\cS^e$, hence, restricting the range of scheduling choices. Actually, the introduction of cross-traffic is shown to have significant implications on the stability behavior of the scheduling policies studied later.

The sets $\cS^e, e\in\cE$ may be overlapping, that is, a service vector may be available under one or more environment states.  Let $\cS=\bigcup_{e\in\cE} \cS^e$. It is assumed that each service vector set $\cS^e,e\in\cE$ is {\em complete}, that is, for each $e\in\cE$ and any $q\in\cQ$
\begin{equation}
\label{propS}
(S_1,S_2,...,S_{q-1},S_q>0,S_{q+1},...,S_Q)\in\cS^e \Rightarrow
(S_1,S_2,...,S_{q-1},S_q=0,S_{q+1},...,S_Q)\in\cS^e.
\end{equation}
Hence, any `sub-vector' of a service vector in $\cS^e$ (i.e. with one or
more {\em positive} components reduced to zero) is also\footnote{Note that if any service vector in $\cS^e$ has no negative components, then the zero vector $(0,0,...,0)$ must be in $\cS^e$ as a sub-vector of the former vector, due to completeness. But if each service vector in $S^e$ has at least one negative component, the zero vector does not necessarily have to be in $\cS^e$ unless it is by design.}
a service vector
%\footnote{The results in this paper hold with a looser definition of
%{\em completeness} for each $\cS^e, e\in\cE$, that is: for each
%$q\in\cQ$, if $S=(S_1, S_2, ..., S_q>0, ..., S_Q) \in\cS^e$, then there
%exists $S'=(S'_1,  S'_2, ..., S'_q, ..., S'_Q) \in\cS^e$, such that
%$S'_q =0$ and $S'_{q'}\leq S'_{q'}$ for all $q'\in\cQ-{q}$.}
in $\cS^e$. The  reason for requiring {\em completeness} of each $\cS^e$ is to accommodate the following situation: {\em when some queues become empty and ceases receiving service, the resulting effective service vector is a feasible one.} Under the latter perspective, the imposed
assumption (\ref{propS}) is a natural one indeed. As seen below, it allows us to naturally handle schedules which provide zero service rate to empty queues.

The key issue is choosing the service vector $S(t)\in\cS^{e(t)}$ at time $t$, when the environment is in state $e(t)$ and the vectors $\cS^{e(t)}$ are available to choose from. In general, the decision can be based on the observable histories of the workload $\{X(z),z\leq t\}$, the environment $\{e(z),z\leq t\}$, and prior service choices $\{S(z), z<t\}$. The scheduling policy is the overall trace of service vector choices $\bS=\{S(t), t\geq 0\}$. Our primary objective is to design schedules $\bS$ which maximize the system throughput (keep the system stable under the maximum possible load $\rho$), while being {\em robust} and utilizing minimum information, like the current workload and environment states, with no knowledge of the actual load $\rho$ and the environment time proportions $\{\pi^e,e\in\cE\}$. We elaborate on such issues later.

We are interested in natural schedules $\bS=\{S(t),
t\geq 0\}$ that never apply positive service to empty queues. That is, whenever $X_q(t)=0$ the scheduler chooses a  service vector $S(t)\in\cS^{e(t)}$ with $S_q(t)\leq 0$. This is possible because we have assumed that the sets $\cS^e, e\in\cE$ are {\em complete}. Therefore, we can write
\eq
\label{nonidling}
X(t)=X(0)+\int_0^tA(z)dz - \int_0^tS(z)dz
\en
without having to explicitly `compensate' for any idling time.

\section{The Stability Issue}
\label{secStability}

In the interest of robustness of the results, we employ the `lightest' possible (see below) concept of stability, that is, {\em rate stability} \cite{BW:93}. Specifically, we call the system stable iff
\eq
\label{eqRate}
\lim_{t\go\infty}\frac{X(t)}{t}=
\lim_{t\go\infty}\(\frac{X_1(t)}{t},\frac{X_2(t)}{t},...,
\frac{X_q(t)}{t},...,\frac{X_Q(t)}{t}\)= 0.
\en
Note that from (\ref{nonidling}) and (\ref{eqratestab}), rate-stability implies that $\rho=\lim_{t\go\infty} \{\int_0^tS(z)dz /t\}$. Moreover, when the traffic trace involves pure `job-arrivals' ($\delta$-jumps) with zero workload arrival rate between them, then
rate-stability  (\ref{eqRate}) implies that the long-term job departure rate from each queue is equal to the long-term job arrival rate \cite{ArmB:03}. Therefore, there is {\em flow conservation} through the system and the inflow at each queue is equal to the outflow. On the contrary, when the system is unstable there is a inflow-to-outflow deficit, which accumulates in the queues. This is consistent with engineering intuition and, in that sense, the concept of rate-stability is quite natural.
Of course, it can be further tightened by imposing progressively heavier statistical assumptions on the traffic and environment traces. We resist doing that at this point, in order to preserve the generality of the results and keep them as robust and `assumptions-agnostic' as possible.

\noindent
\begin{Definition} [Stability Region]\rm
We define formally the stability region $\cR$ of the system as the set of traffic loads $\rho\in\reals^Q_{0+}$ for which there exists a scheduling policy $\bS=\{S(t),t\geq 0\}$ under which the system is rate-stable (\ref{eqRate}) for {\em all} traffic traces $\bA=\{A(t), t\geq 0\}$ with $\rho(\bA)=\rho$ and {\em all} environment traces  $\bE=\{e(t), t\geq 0\}$ with $\pi^e(\bE) = \pi^e, e\in\cE$.

As shown below, the universal stability region $\cR$ can be characterized as
\begin{multline}
\label{defstab1}
\cR(\cS^e,\pi^e,e\in\cE) = \\
\left\{ \rho \in \reals_+^Q : 0 \leq \rho \leq
\sum_{e\in\cE} \pi^e \sum_{S \in\cS^e}\phi^e_S S,
\mbox{ for some } \phi^e_S \geq 0  \mbox{ with }
\sum_{S \in \cS^e} \phi^e_S = 1, \ e\in\cE \right\}
\end{multline}
\end{Definition}

The intuition is that $\rho$ is in the stability region $\cR$ if it is dominated (covered) by a convex combination of the service vectors $S\in\cS$, induced under the various service vectors in $\cS^e, e\in\cE$. Thus, $\cR$ is the `weighted sum' of the various `stability
regions' generated by the individual sets $\cS^e$ for each state $e\in\cE$ of the environment.

If $\rho\in\cR$ and $\pi^e, e\in\cE$ were known in advance and $\phi^e_S$ could be computed, then selecting each mode $S\in\cS^e$ for a fraction $\phi^e_S$ of the time while the system is in environment state $e\in\cE$ would keep the system stable. This could be achieved through round-robin or randomized algorithms. A scheduling algorithm which maintains stability (\ref{eqRate}) for any $\rho \in\cR$ is referred to as {\em throughput maximizing}.
However, we are primarily interested in adaptive scheduling schemes which maintain stability (\ref{defstab1}) for all $\rho\in\cR$, without actual prior knowledge of $\rho$ or $\pi^e$. The {\em cone schedules} defined below are shown to provide such universal stability for any traffic load in $\cR$, while being agnostic to particulars of the traffic and environment traces $\rho(\bA)$ and $\pi^e(\bE) = \pi^e, e\in\cE$; they respond only to current workload and environment state.

In general, the stability behavior of scheduling rules could require the arrival  trace to satisfy stronger conditions than those above. For example, restricting the study to  Markovian or stationary arrival processes, or disallowing mixing, may provide special cases of stability. 
Instead, we allow the arrival traffic trace $\bA$ and environment trace $\bE$ to be designed by an adversary to stress the system. Consider for example an arrival trace where arrivals to queue $q$ are deliberately correlated to the environment states when $q$ cannot be served at maximum capacity. Even further, an adversarial trace may push arrivals to queues in a state-dependent way which responds to the scheduling rules themselves. These are very difficult to capture by a natural probabilistic framework, but are simply treated as  particular traffic traces here.

To motivate the definition of the stability region for the processing system under consideration, we examine first the case where $S_q\geq 0$ for all $q\in\cQ$ and there is only one environment state ($E=1$, no environment fluctuation); that is, service is always non-negative and all service vectors $\cS$ are available at every point in time. Under the trace-based perspective employed in this paper, it is known \cite{ArmB:03} that for any load $\rho$ in the region
\eq
\nonumber
\left\{\rho \in \reals_{0+}^Q :~
\rho \leq \sum_{S \in \cS}\phi_S S ~\mbox{for some }
\phi_S \geq 0, \ S \in \cS, ~ \mbox{ such that }
\sum_{S \in \cS} \phi^S\leq 1 \right\}
\en
the system can be made rate stable with an appropriate scheduling rule. The non-negative parameters $\phi_S, S \in \cS$ are essentially proportional weights, which are chosen so that the load vector $\rho$ is component-wise dominated by the weighted linear combination $\sum_{S \in \cS}\phi_S S$ of the service vectors.

Extending this `geometric' stability perspective to allow cross traffic and varying environment states is not a trivial task. Intuition may suggest that the stability region in networks in fluctuating environments should be reduced according to how
often each mode is available.  Consider the following simple network to illustrate that the distribution of environment states $\{\pi^e, e\in\cE\}$ is critical to stability.  Take a 2-queue network with three service vectors, $S^1=(1,0), S^2=(0,1), S^3=(1,1)$.  Clearly, if
all vectors are available all the time, by employing always $S^3$ the system can accommodate any input vector $(\rho_1,\rho_2)\in [0,1]^2$.
On the other hand, if there are two environment states $\cE=\{e_1,e_2\}$ with service vector sets $\cS^{e_1}=\{S^1,S^2\}$ and $\cS^{e_2}=\{S^3\}$ with $\pi^{e_1}=0.5,~\pi^{e_2}=0.5$, 
 then the system can accommodate
any input vector $\rho \geq 0$ satisfying the conditions $\rho_1 + \rho_2 \leq 1.5, \rho_1 \leq 1$, and $ \rho_2 \leq 1$. 

\begin{figure}
\begin{center}
\includegraphics[width=0.3\linewidth]{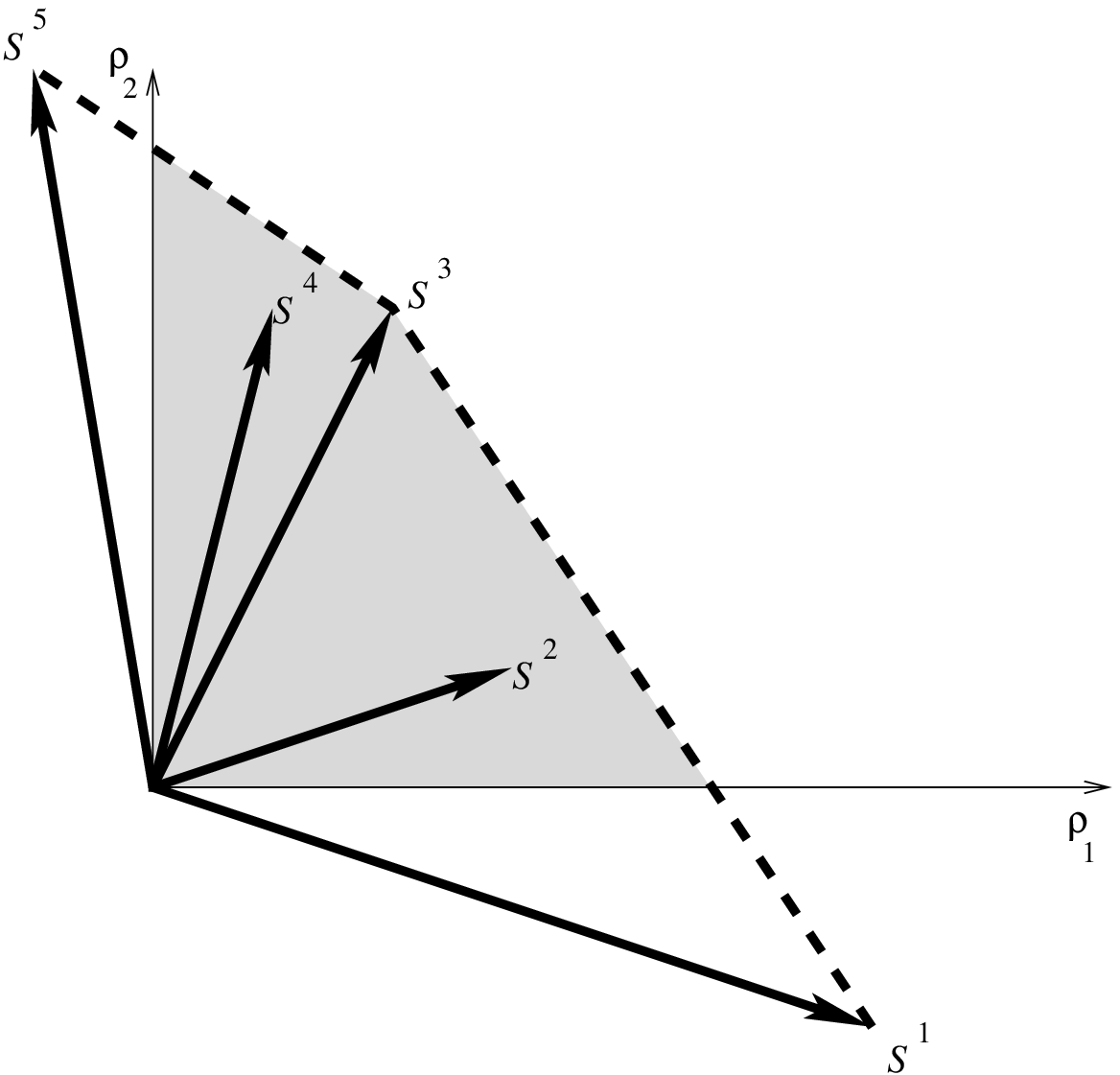}
\includegraphics[width=0.3\linewidth]{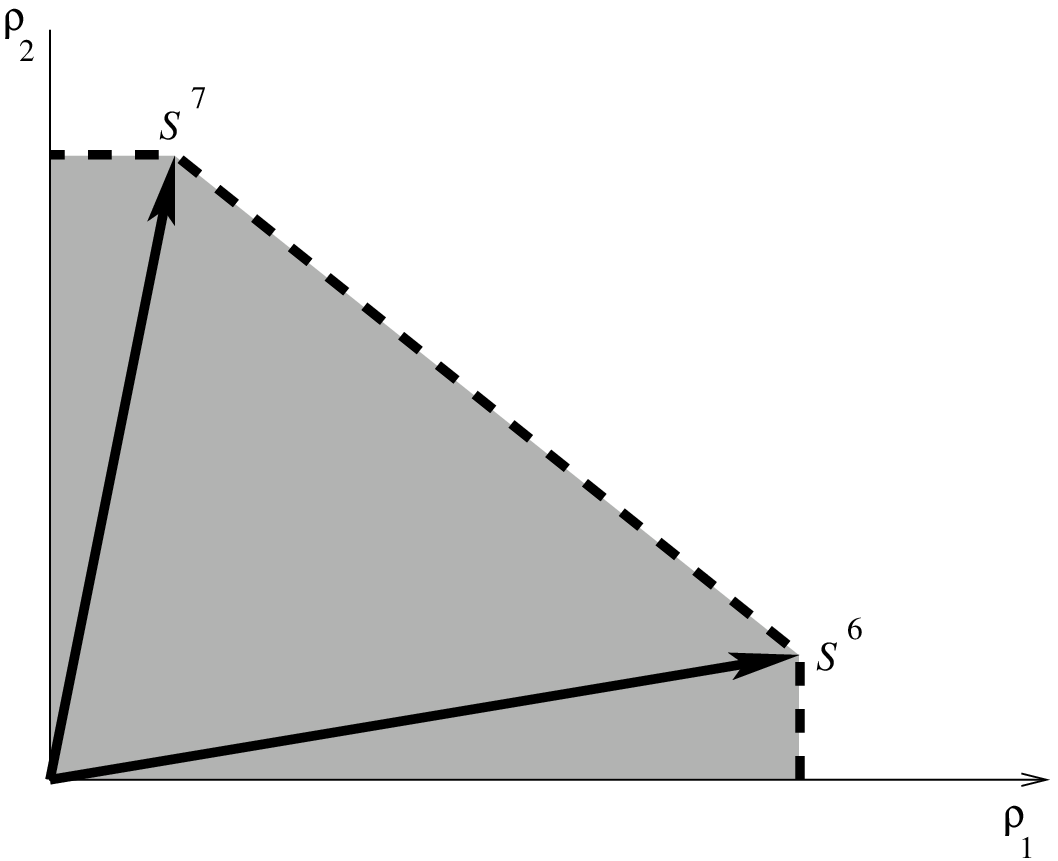}
\includegraphics[width=0.3\linewidth]{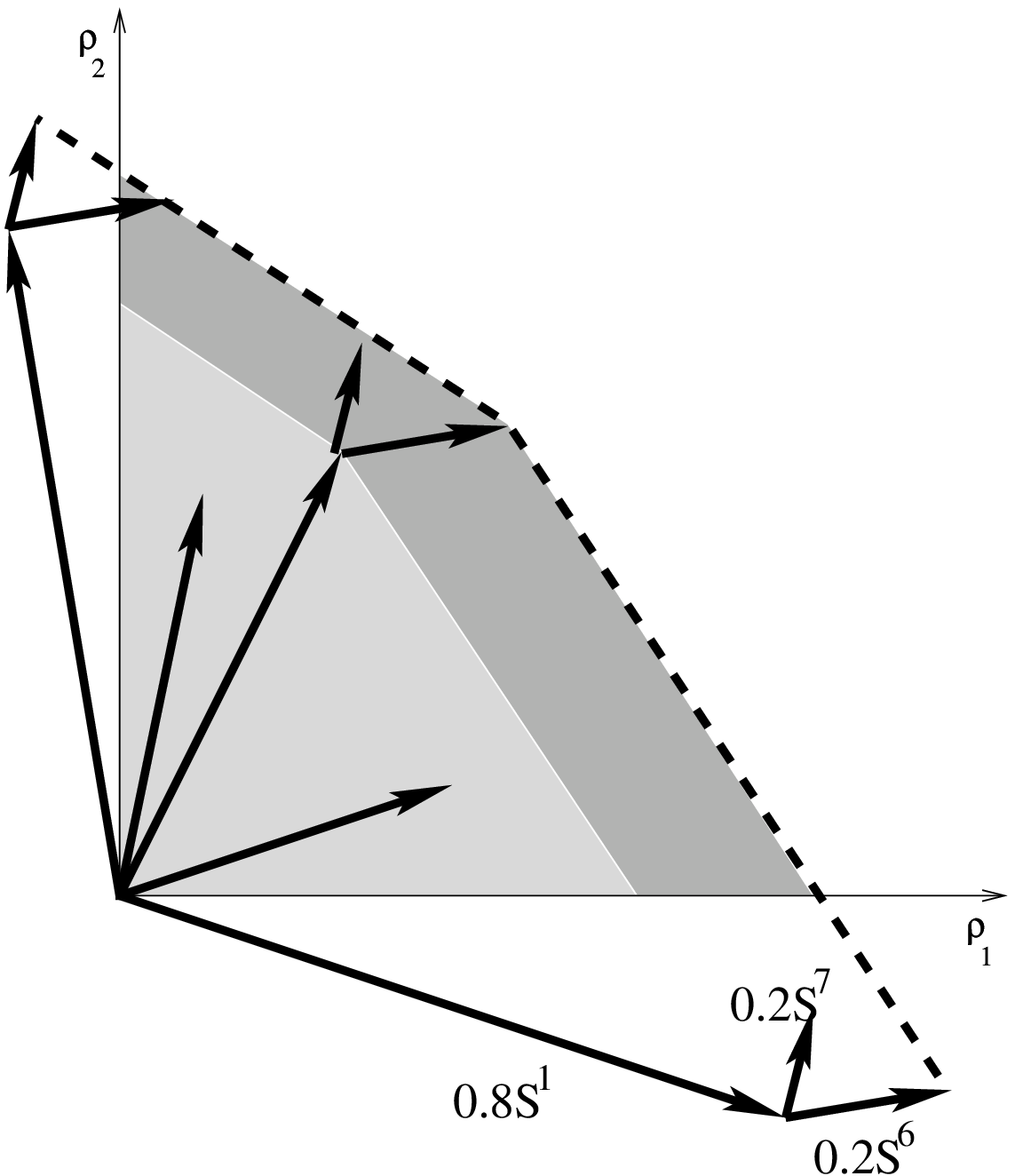}
\caption{{\bf The stability region. } The set of allowable arrival
rate vectors $\rho$ is called the stability region $\cR$.  Two
separate sets of service vectors are shown in the first two
plots, with their respective stability regions if they were the only environment state, and available $100\%$ of the time.  The third plot shows the stability region when $\pi_1=0.8$ and $\pi_2=0.2$. This corresponds to the environment state fluctuating so that $80\%$ of the time, the service vectors from the first group are available, and $20\%$ of the time the service vectors from the second group are available to be scheduled. For any $\rho$ in the region
$\cR$ above, there is a convex combination of service modes within
the resource sets which would apply a total service rate to each
queue which is at least the arrival rate to that queue. For $\rho$
outside $\cR$ there is no such combination.   Service modes $S^2$ and $S^4$ are strictly dominated by a convex combination of other service vectors and therefore do not contribute to the stability region (and in fact need not be utilized to maintain stability).
Service vectors with 
negative components such as $S^1$ and $S^5$ above may contribute to
the stability region without being inside the stability region
itself. The stability region for the combination of environments can be seen to be the weighted sum of the two original stability regions, with care taken to the impact of extreme points with negative components. }
\label{figStabReg}
\end{center}
\end{figure}

However, a different configuration of the service vector sets, say $\cS^{e_1}=\{S^1\}$ and $\cS^{e_2}=\{S^2,~S^3\}$ with $\pi^{e_1}=0.5, \pi^{e_2}=0.5$, yields $\rho_1\in [0,1]$ and $\rho_2\in [0,0.5]$ for stability. Note that although the sets $\cS^{e_1}$ and $\cS^{e_2}$ ensure that each service vector is available for the same portion of time in both scenarios, the relative combinations of the available service vectors change the stability region. We illustrate (and generalize) this perspective in Figure \ref{figStabReg}.

We establish first that if $\rho\notin\cR$, it is impossible to maintain stability and flow conservation in all queues, no matter what scheduling policy one employs. At least one queue will suffer an outflow deficit
(compared to its inflow), which will accumulate in the queue and cause its workload to explode linearly it in time.

\bProposition[Instability]\rm
\label{propUnstable}
For any arbitrarily fixed traffic traffic trace $\bA$ and environment trace $\bE$, we have
\eq
\label{equnstable}
\rho(\bA)\notin\cR\Longrightarrow  \limsup_{t\go\infty} \frac{X_q(t)}{t}>0 ,
\en
for at least one queue $q\in\cQ$ under {\em any} scheduling policy.
\eProposition

\bProof
For convenience, we drop the fixed argument $\bA$ from $\rho(\bA)$ and write it traffic load as simply $\rho$, and proceed by contradiction. If (\ref{equnstable}) does not hold, then  from (\ref{nonidling}) we must have $\lim_{t \go \infty} \frac{\int_0^tS(z)dz}{t}=\lim_{t \go \infty} \frac{\int_0^tA(z)dz}{t}+\lim_{t\go\infty}\frac{1}{t}X(0)=\rho$. But then we have

\begin{align*}
 \rho &= \lim_{t \go \infty} \frac{\int_0^tS(z)dz}{t} \\
&= \lim_{t \go \infty} \frac{ \int_0^t \sum_{e\in\cE}\sum_{S\in\cS^e}\bI_{e(z)=e,S(z)=S} S dz}{t} \\
&= \sum_{e\in\cE} \pi^e \sum_{S \in\cS^e}\hat{\phi}^e_S S
\end{align*}

where $\hat{\phi} = \lim_{t \go \infty} \frac{ \int_0^t \bI_{S(z)=S} S dz}{t}$ satisfies $\hat{\phi}^e_S \geq 0$ and $\sum_{S \in \cS^e} \hat{\phi}^e_S = 1$, which satisfies (\ref{defstab1}).
We then easily get (arguing by contradiction) that $\limsup_{t\go\infty} X^q(t)/t>0$ for at least one queue $q\in \cQ$.
\eProof

\section{Cone Schedules and their Geometry}
\label{secSchedules}

We focus in this paper on schedules that are workload-aware and resource-aware but not
rate-aware; that is, the system's operator can observe and respond to both the environment state $e(t)$ and the workload state $X(t)$, but has no
knowledge of the long-term load vector $\rho$ and state probabilities $\pi^e$.

In particular, we examine a family of resource allocation policies that are called Cone Schedules and are parameterized by a fixed matrix
$\bB$.  These schedules select the service vector $\hat{S}\in\cS^{e(t)}$ that has the maximal projection on $\bB X(t)$, when the workload state is $X(t)$ and the environment state is $e(t)\in\cE$. Specifically:

\begin{Definition}[Cone Schedules]\rm
Given a fixed $Q\times Q$ real matrix $\bB$, a {\em cone schedule} is one that, when the environment is in state $e\in\cE$ and the the workload is $X\in\reals_{0+}^Q$, it selects a service vector $\hat{S}^e(X)$ in the set
\eq
\label{cone}
\hat{\cS}^e(X) =
\arg\max_{S\in\cS^e} \< S,\bB X \> =
\{\hat{S}\in\cS^e : \< \hat{S},\bB X \> =
\max_{S\in\cS^e} \< S,\bB X \> \} 
\en
 which satisfies $S_q=0$ whenever $S_q=0$. We show that such a vector must be contained in $\hat{\cS}^e$ by proposition \ref{norate} below.
The set $\hat{\cS}^e(X)\subseteq \cS^e$ is nonempty, but may contain several service vectors in $\cS^e$, in which case one is arbitrarily chosen by the cone schedule. Note that
\eq
\label{cone2}
\< \hat{S}^e(X),\bB X \> = \max_{S\in\cS^e} \< S,\bB X \>,
\en
so the chosen $\hat{S}^e(X)$ is one of maximal projection on $\bB X$ amongst those in $\cS^e$. Therefore, the service vector $\hat{S}(t)$ chosen by the cone schedule at time $t\geq 0$ is
\eq
\nonumber
\label{cone3}
\hat{S}(t) \in \hat{\cS}^{e(t)}(X(t)) =
\arg\max_{S\in\cS^{e(t)}} \< S,\bB X(t) \>,
\en
based on the observed current workload $X(t)$ and environment state $e(t)$.
\end{Definition} 

Notice that the maximization $\<S, \bB X \> = \sum_q S_q (\bB X)_q$ ensures that cone schedules follow some important intuition for a scheduling rule. We see that $(\bB X)_q$ is increasing in $X_q$ and decreasing in $X_p$ for $p \neq q$. This will increase whenever $X_q$ comes to dominate other queues. By maximizing this sum, the cone schedules all prefer large positive service rates $S_q$ whenever $(\bB X)_q$ is large and positive. Thus the schedules will prefer remove the most workload from the longer queues, and restrict the cross-traffic added to those longer queues. The relationship to performance and load balancing is discussed in section \ref{secPerf}.

\begin{Proposition}[Matrices $\bB$ with Negative or Zero Off-Diagonal Elements]\rm
\label{norate}

If the cone schedule matrix $\bB=\{B_{ij}, i,j\in\cQ\}$  has {\em negative or zero off-diagonal elements} ($\bB_{ij}\leq 0, i\neq j$) and the service vector sets $\cS^e$ are {\em complete} for each environments state $e\in\cE$, then there must exist some  $\hat{S}^e(X)\in\hat{\cS}^e(X)$ for which we have
\eq
\nonumber
X_q = 0 \implies \hat{S}^e_q (X)\leq 0
\en
for each $q\in\cQ$. Thus, for such $\bB$ matrices, the corresponding cone schedules can always select service vectors that provide {\em no positive service rate to an empty queue}.

\end{Proposition}

\bProof
Given a workload vector $X$ such that $X_q=0$ for some (empty) queue $q\in\cQ$, let us examine the inner product maximized by the cone schedule (\ref{cone}) in selecting $\hat{S}^e(X)\in\hat{\cS}^e(X)$, that is,
\eq
\label{mxip}
\< S,\bB X\> = \sum_{i\in\cQ}\sum_{j\in\cQ} S_i B_{ij} X_j =
\sum_{i\in\cQ} \{ S_i B_{ii} X_i + S_i ( \sum_{j\in\cQ-\{i\}}
B_{ij} X_j ) \} .
\en
with $S\in\cS^e$. Consider the term corresponding to the empty queue $q$ in the above sum, that is,
\eq
\label{ngbx1}
S_q B_{qq} X_q + S_q ( \sum_{j\in\cQ-\{q\}} B_{qj} X_j ) .
\en
Since $X_q=0$, the first term above is automatically zero, irrespectively of $S_q$ and $B_{qq}$. However, since $B_{qj}\leq 0$ and $X_j\geq 0$ for each $j\in\cQ-\{q\}$, we see that
\eq
\label{ngbx2}
\sum_{j\in\cQ-\{q\}} B_{qj} X_j \leq 0 .
\en

Arguing by contradiction, assume that  $(S_1, S_2,  ..., S_{q-1}, S_q>0, S_{q+1}, ..., S_Q) \in\cS^e$ maximizes (\ref{mxip}) with $S_q>0$. But because $\cS^e$ is assumed to be {\em complete} (\ref{propS}), the vector $(S_1, S_2, ..., S_{q-1}, S_q=0, S_{q+1}, ..., S_Q)$ also belongs to $\cS^e$ and has $S_q=0$, hence, leads to an equal or greater value of (\ref{mxip}) because of (\ref{ngbx1}) and (\ref{ngbx2}).  This establishes a contradiction and implies that the set of service vectors $S$ that maximize (\ref{mxip})  must always include one where $S_q\leq 0$ (provide no positive service rate) for each empty queue $q\in\cQ$ (that is, with workload $X_q=0$).

\eProof
%
%{\red [We should say here that we restrict the following discussion to the interesting/natural case of cone schedules generated by $\bB$ matrices that have negative or zero off-diagonal elements, which do not provide positive service rate to any empty queue.]}

To justify the term `cone' schedule consider the following perspective. Define first the set of workloads $X$ for which the cone schedule would choose the service vector $S$ when the environment is in state $e$, that is:

\eq
\nonumber
\cC_S^e = \left\{X \in \reals_{0+}^Q :
\< S,\bB X \>  =  \max_{S'\in\cS^e} \< S',\bB X \> \right\}
\en

for $S\in\cS^e, e\in\cE$. This is simply the set of workloads $X$ that have maximum projection on $S\in\cS^e$ amongst all other sets in $\cS^e$. Note that $\cC^e_S$ is a {\em geometric cone} because $\< S,\bB X\> \geq \< S',\bB X\>$ implies that $\< S,\bB\alpha X\> \geq \< S',\bB \alpha X\>$ for any positive scalar $\alpha\in\reals_+$ and $S,S'\in\cS^e$. Thus, if $X$ belongs to $\cC_S^e$ then any up/down-scaling $\alpha X$ also belongs to it.

For each environment state $e\in\cE$, the cones $\cC^e_S, S\in\cS^e$ form a {\em partition} of the workload space, that is,

\eq
\nonumber
\bigcup_{S \in \cS^e} \cC_S^e = \reals_{0+}^Q.
\en

In general, some cones may actually be degenerate (like those
corresponding to service vectors in $\cS^e$ that are fully dominated component-wise by others in $\cS^e$) and several cones may
share common boundaries. Observe that the cone schedule can now be
geometrically defined as follows:
\eq
\nonumber
\boxed{
\mbox{When the environment state is } e
\mbox{ and the workload}\ X \in \cC_S^e \implies
\ \mbox{choose } \ \hat{S}^e(X)=S \in \cS^e .}
\en
The cone structure of the sets $\cC_S^e$ motivates the name {\em cone schedules}.

\begin{figure}[!h]
\begin{center}
\includegraphics[width=0.5\linewidth]{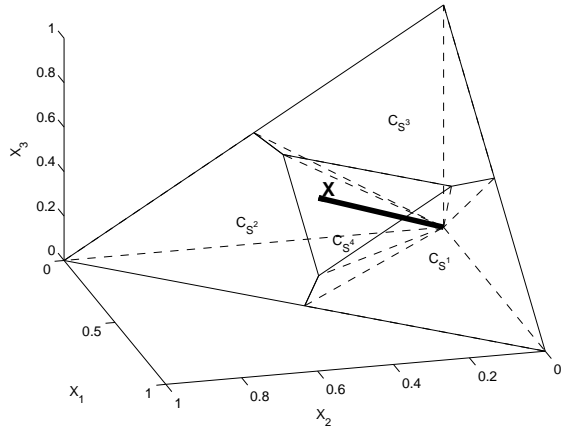}
\caption{The cone schedules assign a service vector from $\cS^e$ by identifying the location of $X$ with respect to the cones formed by $\cC_S^e$. This figure shows the cone structure for a system with $Q=3$ queues and 4 service vectors for this particular environment.
When $X$ is in cone $\cC^e_S$, then service vector $S$ corresponding to that cone is used. The vector $X$ will fluctuate within $\Re^3$, switching between service vectors when the arrivals and departures cause $X(t)$ to cross a cone boundary, or when the environment state changes. The cone boundaries are influenced by the environment state and the matrix $\bB$.
}
\label{fig3dCone}
\end{center}
\end{figure}

When the environment is in state $e\in\cE$ and the workload $X$ is in the interior of the non-degenerate cone $\cC_S^e$, then the only service vector that can be used by the cone schedule is $S\in\cS^e$. However, if $X$ is on the boundary of several adjacent cones (for example,
$X\in\cC^e_{S^1}\bigcap\cC^e_{S^2}\bigcap\cC^e_{S^3}$), then any of
the service vectors corresponding to these cones can be used ($S^1$,
or $S^2$, or $S^3$). Therefore,
given a workload vector $X$, we want to define the cone it belongs
to, which consequently specifies what service vector the cone schedule
ought to use. We proceed in this direction below.

To take another perspective, recall that when the environment is in state $e\in\cE$ and the workload is $X$, then the cone schedule chooses a service vector $\hat{\cS}^e(X)$ in the set
\eq
\nonumber
\hat{\cS}^e(X)=\left\{ \hat{S} \in\cS^e : \<\hat{S},\bB X \> =
\max_{S \in \cS^e} \< S ,\bB X \> \right\}\subseteq \cS^e;
\en
any vector is arbitrarily chosen, if there are more than one vector in the set $\hat{\cS}^e(X)$. When $X$ is in the interior of the (non-degenerate) cone $\cC^e_S$, then $\hat{\cS}^e(X)=\{S\}$ is a singleton and $\hat{S}^e(X)=S$. 
This follows since the interior of a cone denotes all workload vectors $X$ for which the inner product $\< S ,\bB X \>$ is {\em uniquely} maximized by $S$.

\begin{figure}[!h]
\begin{center}
\includegraphics[width=0.5\linewidth]{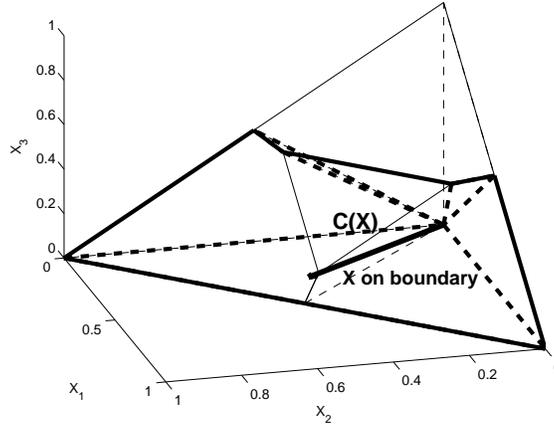}
\caption{For workload vectors $X$ which lie precisely on the boundary of two or more cones,  the cone $\cC^e(X)$ is the union of all of the cones in $\cC^e$ which include $X$. In contrast to Fig. \ref{fig3dCone}, where $X$ was interior to a single cone, the above illustration shows $X$ at the boundary of 3 of the $\cC_S$ cones. In this case $\cC(X)=\cC_{S^1} \cup \cC_{S^2} \cup \cC_{S^4}$ includes all the elements of the three different cones. This definition is important in the proof because it captures the workload vectors which share an optimal service vector with $X$.}
\label{figConeBoundary}
\end{center}
\end{figure}

To cover the general case of $X$ being on a cone boundary (perhaps, a common boundary of several cones), we define the `surrounding' cone of the workload vector $X$ as
\eq
\nonumber
\cC^e(X)=\bigcup_{S \in \hat{\cS}^e(X)}\cC^e_S
\en
For example, if $X$ is on the boundary of $\cC^e_{S^1}$ and
$\cC^e_{S^2}$ only, then $\cC^e(X)= \cC^e_{S^1} \bigcup \cC^e_{S^2}$.  Note that the above definitions lead to the following equivalence
\eq
\nonumber
\cC^e(X) \subseteq C^e(Y) \Leftrightarrow \hat{\cS}^e(X) \subseteq
\hat{\cS}^e(Y) ,
\en
as well as
\eq
\nonumber
\cC^e(X) \subseteq \cC^e(Y) \Rightarrow X \in \cC^e(Y)
\en
for any two workload vectors $X,Y\in\reals_{0+}^Q$ and environment state $e\in\cE$.  This is illustrated in Fig. \ref{figConeBoundary}.

Note that if $Y \in \cC^e(X)$ then there must exist a service vector $\hat{S} \in \cS^e$ for which both $\<S, \bB X\>$ is maximized at $\hat{S}$ and $\<S, \bB Y\>$ is  maximized at $\hat{S}$, and if $Y \notin \cC^e(X)$ then no such vector can exist.

We observe that $X$ cannot be on an interior boundary of $\cC^e(X)$ (the only boundary it could be on is where the cone meets an axis because of the non-negativity constraint). If $X$ were on an interior boundary then there must exist a direction vector $\delta \neq 0$ for which $(X+\lambda\delta) \geq 0$ and $(X+\lambda\delta) \notin \cC^e(X)$ for an arbitrarily small positive scalar $\lambda$. This means that there exists some service vector $S^\delta \in \cS^e$ for which $\<S^\delta, \bB (X+\lambda\delta )\> > \<S, \bB (X+\lambda\delta )\>$ for all $S \in \hat{\cS}^e$. But since $S^\delta \notin \hat{\cS}^e(X)$ we also have $\<S^\delta, \bB X \> < \<S, \bB X\>$ for all $S \in \hat{\cS}^e$.  This leads to the inequality
\eq
\nonumber
\lambda ( \<S^\delta , \bB \delta \> - \<S, \bB \delta \> ) > \< S, \bB X \> - \< S^\delta \bB X \> > 0.
\en

Since the left hand side can be  made arbitrarily small this leads directly to a contradiction and we conclude that $X$ is indeed on the strict interior of $\cC^e(X)$. This observation becomes critical in the proof of stability.

\begin{figure}
\label{ConeEnv}
\begin{center}
\includegraphics[width=0.5\linewidth]{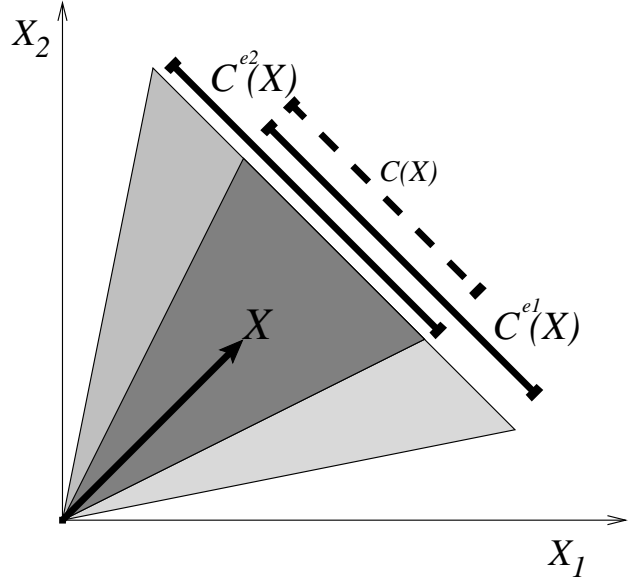}
\caption{The cone $\cC(X)$ over environments $\cE$ is illustrated. Here, $X$ is in the cones $\cC^{e_1}(X)$ and  $\cC^{e_2}(X)$ for the two environments $e_1$ and $e_2$. The cone $\cC(X)$ is the intersection of both of those cones. Since $X$ is known to be on the interior of each cone, $X$ is also on the interior of $\cC(X)$.  }
\label{figConeEnv}
\end{center}
\end{figure}

Finally, we define the cone around $X$ with respect to {\em all} environment states $e\in\cE$ as

\eq
\nonumber
\cC(X) = \bigcap_{e\in\cE} \cC^e(X).
\en
The cone $\cC(X)$ is illustrated in Fig \ref{figConeEnv} is of course non-empty because $X$ belongs to each cones $\cC^e(X), e\in\cE$. This is the cone of workloads $Y$ for which, at each environment state $e\in\cE$, the cone schedule could have selected for $Y$ the same service vector as for $X$ (fixed), that is,

\eq
\nonumber
\cC(X)=\{Y\in\reals^Q_{0+} :
\hat{\cS}^e(Y)\bigcap\hat{\cS}^e(X)\neq \emptyset,
\mbox{ for each } e\in\cE\} ;
\en

hence, when $Y\in\cC(X)$, then for each $e\in\cE$ we have  $\hat{S}^e(Y)\in\hat{\cS}^e(X)$, besides $\hat{S}^e(X)\in\hat{\cS}^e(X)$ of course. Viewed another way,

\eq
\nonumber
\cC(X)=\{Y\in\reals^Q_{0+} :
\<\hat{S}^e(Y),\bB X\> =
\max_{S\in\cS^e}\<S,\bB X\>  \mbox{ for each } e\in\cE\} ,
\en

that is, when $Y\in\cC(X)$, then for each $e\in\cE$ we have that $\hat{S}^e(Y)$ has maximal projection on $\bB X$, besides also having maximal projection on $\bB Y$ (by definition).

 We note that since $X$ is strictly on the interior of each cone $\cC^e(X)$ and there are finitely many environment states in $\bE$ then $X$ is strictly on the interior of $\cC(X)$.

The cone $\cC(X)$ turns out to be of key importance in the stability proof below. This completes the geometric picture of cone schedules. 
%
%{\red [{\blue KEVIN:} We definitely need a nice 2-dim or 3-dim graph showing the various cones discussed in this section. Actually, it should show the intersections of the 3-dim cones with a cutting plane.]}

\section{Universal Stability of Cone Schedules}
\label{secProof}

We are primarily interested in the throughput maximizing properties of cone schedules for various families of matrices $\bB$, given the traffic load $\rho$. The following theorem establishes that stability can be maintained for any $\rho\in\cR$ by rich families of matrices $\bB$.

Consider a cone schedule generated by the matrix $\bB$ and operating on any arbitrarily fixed system $\Sigma$ chosen from the class $\mathfrak{S}$ of processing systems defined by:
\begin{enumerate}
\item
some set of queues $\cQ$ and some set of environment states $\cE$,
\item
some environment trace $\bE \in \mathfrak{E}(\pi^e, e\in\cE)$, as per (\ref{envtrst}),
\item
some (non-empty) service vector sets $\cS^e, e\in\cE$ that are {\em complete}, as per (\ref{propS}),
\item
some traffic trace $\bA=\{A(t), t\geq 0\}\in\mathfrak{A}(\rho)$ with load $\rho(\bA)=\rho$, as per (\ref{trtrst}).
\end{enumerate}

%Define further the system class $\mathfrak{S}^+$ as the subset of systems in $\mathfrak{S}$ which additionally satisfy $S_q \geq 0$ {\em for all} $q\in\cQ, S\in\cS^e,e\in\cE$. And define the system class $\mathfrak{S}_-$ as the subset of systems in $\mathfrak{S}$ which additionally satisfy $S_q < 0$ {\em for some} $q\in\cQ, S\in\cS^e,e\in\cE$.

\begin{Theorem}[Universal Stability of Cone Schedules]
\rm
\label{thrate}
Given the above assumptions if $\bB$ is positive-definite, symmetric and has negative or zero off-diagonal elements ($B_{qp}\leq 0, p\neq q\in\cQ$), then
\eq
\rho(\bA)\in\cR(\cS^e,\pi^e,e\in\cE) \implies \lim_{t\go\infty}\frac{X(t)}{t}=0
\en
{\em universally} on $\mathfrak{S}$ . That is, each system in $\mathfrak{S}$ is (rate) stable under such a cone schedule, when $\rho(\bA)\in\cR(\cS^e,\pi^e,e\in\cE)$.

\end{Theorem}

It turns out that $\bB$ being positive definite and having nonpositive off-diagonal elements are both necessary for universal stability, which was shown in \cite{RoB:09}. 

To see why nonpositive off diagonal elements are required, consider a simple network with $Q=2$ queues and $E=1$ environment state, where $\bB=[ 2, 1; 1, 2]$ is used. If $S^1=(1, 0)$ and $S^2=(0,3)$ are the  two available service vectors then $\<S^1,\bB X \> = 2X_1+X_2$, and $\<S^2,\bB X \> = 3X_1+6X_2$. Since $\<S^2,\bB X \>$ strictly dominates $\<S^2,\bB X \>$ for any nonzero workload, $S^1$ would never be selected and any arrival process with $\rho_1>0$ will be unstable.

To see why positive definiteness is required, consider a simple network with $Q=2$ queues and $E=1$ environment state, where $\bB=[ 1, -2; -2, 1]$ is used. Let $S^1=(1, 1), S^2=(0,0), S^3=(1,0)$ and $S^4=(0,1)$ be the available service vectors. Then we have 
 $\<S^1,\bB X \> = -X_1-X_2<0=\<S^2,\bB X \>$, and $S^1$ will never be selected. The effective service rates applied to the queues $\hat{S}_q=\lim_{t\go\infty}\frac{1}{t}\int_o^tS_q(z)$  must then satisfy $\hat{S}_1+\hat{S}_2 \leq 1$. Now $\rho = (\rho_1, \rho_2)$ with $0.5<\rho_q \leq 1$ is contained within $\cR$ by (\ref{defstab1}), but rate stability cannot possibly be achieved in (\ref{nonidling}). The parameters of the non-positive-definite matrix $\bB$ cause the cone schedule to avoid utilizing $S^1$, which is critical for rate stability because it lies on the convex hull of $\cR$.

\subsection{Proof of Theorem \ref{thrate} }
\label{prf}

We prove rate stability via a sequence of intermediate steps.

Consider any arbitrarily fixed environment trace $\bE=\{e(t),t\geq0\}$, such that $\cS^e$ is {\em complete} and
\eq
\nonumber
\lim_{t\go\infty} \frac{\int_0^t \ind{e(z)=e}dz}{t}=\pi^e
\en
for each $e\in\cE$. Consider also any arbitrarily fixed traffic trace $\bA=\{A(t),t\geq 0\}$ satisfying
\eq
\nonumber
\lim_{t\go\infty} \frac{\int_0^t A(z)dz}{t}=\rho(\bA)\in\cR(\cS^e,\pi^e,e\in\cE).
\en

We note that while $\bA$ and $\bE$ are fixed, they can be generated arbitrarily, including by an underlying stochastic process or an adversary. Recall that by Proposition (\ref{norate}) when $\bB$ has negative or zero off-diagonal elements the generated cone schedule applies no positive rate to empty queues. Therefore,
\eq
\label{xaxaxa}
X(t)=X(0)+\int_0^tA(z)dz - \int_0^t S(z)dz
\en
for the workload $X(t)$ at time $t$ -- as in (\ref{nonidling}) --
without having to compensate for any idle time.

\bProposition\rm
Under the conditions conditions of Theorem \ref{thrate}, the service vectors $\hat{S}^e(X)\in\hat{\cS}^e(X)$ selected by the cone schedule under various environment states $e\in\cE$ satisfy
\eq
\label{lempolar}
\<\rho, \bB X\> \leq \sum_{e\in\cE} \pi^e \<\hat{S}^e(X), \bB X \> =
\sum_{e\in\cE} \pi^e \max_{S\in\cS^e} \<S, \bB X \>.
\en
for each workload $X\in\reals_{0+}^Q$.
\eProposition

\bProof
First, choose any workload $X$ and fix it.  Since $\rho\in\cR$, we have $\rho\leq\sum_{e\in\cE} \pi^e \sum_{S \in \cS^e}\phi^e_S S$ according to (\ref{defstab1}), or
\eq
\label{dmnt}
 0 \leq  \rho_q\leq\sum_{e\in\cE} \pi^e \sum_{S \in \cS^e}\phi^e_S S_q,
\mbox{ for each } q\in\cQ,
\en
for some positive weights $\phi^e_S \geq 0$ such that $\sum_{S \in \cS^e}\phi^e_S \leq 1$.

We denote $v_q=(\bB X)_q$ and note that this may be negative for some $q\in\cQ$. We examine, the following two cases:
\begin{enumerate}
\item
If $v_q =(\bB X)_q \geq 0$, we get from (\ref{dmnt}) that
\eq
\rho_q v_q \leq
\sum_{e\in\cE} \pi^e \sum_{S \in \cS^e}\phi^e_S S_q v_q .
\en
\item
If $v_q=(\bB X)_q < 0$, we have
\eq
\nonumber
\rho_q v_q \leq 0,
\en
since $\rho_q \geq 0$.
\end{enumerate}
Combining the two cases, we get
\eq
\nonumber
\rho_q v_q \leq
\sum_{e\in\cE} \pi^e \sum_{S \in \cS^e}\phi^e_S S_q\ind{v_qS_q \geq 0} v_q \leq \sum_{e\in\cE} \pi^e \sum_{S \in \cS^e}\phi^e_S S_q\ind{v_q \geq 0} \ind{s_q>0}v_q 
\en

for $q\in\cQ$. Adding the terms up over $q\in\cQ$, we get
\eq
\label{vdom}
\<\rho, v\> \leq
\sum_{e\in\cE} \pi^e \sum_{S \in \cS^e}\phi^e_S \< V(S), v\>
\en
where  $V(S) = (S_q\ind{v_q \geq 0, S_q > 0}),  q\in\cQ)$ is the vector generated by the service vector $S\in\cS^e$ by setting 0 the components  $S_q$ for which $v_q < 0$ and $S_q>0$.

 Now recall that for each $e\in\cE$ and $S\in\cS^e$,  $V(S)$ is a {\em sub-vector} of $S$ (dropping some positive components to 0) and is also in $\cS^e$ because the latter set is {\em complete}. But the service vector $\hat{S}^e(X)$ selected by the cone schedule (\ref{cone2}) has the maximal projection on $v=\bB X$ amongst all those in $\cS^e$, so $\< V(S), v\> \leq \< \hat{S}^e(X), v\>$ for every $S\in\cS^e$. Therefore, (\ref{vdom}) becomes
\eq
\nonumber
\<\rho, v\> \leq
\sum_{e\in\cE} \pi^e \sum_{S \in \cS^e}\phi^e_S \< V(S), v\> \leq
\sum_{e\in\cE} \pi^e \left[ \sum_{S \in \cS^e}\phi^e_S \right] \< \hat{S}^e(X), v\> \leq
\sum_{e\in\cE} \pi^e  \< \hat{S}^e(X), v\>,
\en
where the last inequality holds because $\sum_{S \in \cS^e}\phi^e_S  \leq 1$ for each $e\in\cE$. Putting back $v=\bB X$, we get
\eq
\nonumber
\<\rho, \bB X\> \leq \sum_{e\in\cE} \pi^e  \< \hat{S}^e(X), \bB X\>,
\en
which completes the proof.

\eProof

\begin{lemma} \rm
\label{lemratestab}
We have that $\lim_{t \go \infty}\frac{X(t)}{t} = 0$ implies $\left.
\lim_{t\go \infty} \frac{\int_0^t \hat{S}(z)dz}{t} =\rho \right.$. That
is, the long-term applied service rate is equal to the long-term
traffic load, when the system is (rate) stable.
\end{lemma}
\bProof This is immediately obtained by dividing (\ref{xaxaxa}) by $t$ and letting $t\go\infty$. \eProof

\begin{lemma}\rm
\label{lemwkld}
Consider two arbitrarily fixed, increasing, unbounded time
sequences $\{t_n\}_{n=1}^\infty$ and $\{s_n\}_{n=1}^\infty$ with $s_n\leq t_n$ for each $n\geq 1$.  If
$\lim_{n\go \infty} \frac{t_n-s_n}{t_n}=0$ \ (or equivalently
$\lim_{n \go \infty} \frac{s_n}{t_n}=1$), then
\eq
\nonumber
\lim_{n \rightarrow \infty}  \frac{X(t_n)-X(s_n)}{t_n}=\lim_{n
\rightarrow \infty} \frac{X(t_n)-X(s_n)}{s_n}=0.
\en
\end{lemma}
\bProof
Note that $0\leq\int_{s_n}^{t_n}\ind{\hat{S}(z)=S}dz \leq t_n-s_n$ for
each $S \in \cS=\cup_{e\in\cE}\cS^e$. Dividing by $t_n$ and taking the limit as $n\go\infty$, we get $\left. \lim_{n\go\infty} \frac
{\int_{s_n}^{t_n} \ind{\hat{S}(z)=S}dz}{t_n}=0 \right.$. Recalling that
\eq
\nonumber
X(t_n)-X(s_n)= \int_{s_n}^{t_n}A(z)dz - \sum_{S \in \cS} S
\int_{s_n}^{t_n} \ind{\hat{S}(z)=S}dz,
\en
dividing by $t_n$ and letting $n \go \infty$, we get
\begin{eqnarray}
\nonumber
\lim_{n \go \infty}\frac{X(t_n)-X(s_n)}{t_n}
& = &
\lim_{n  \go \infty} \frac{\int_{s_n}^{t_n}A(z)dz}{t_n} - \lim_{n \go \infty}
\frac{\sum_{S \in \cS} S \int_{s_n}^{t_n}\ind{\hat{S}(z)=S}dz}{t_n} \\
\nonumber
& = &
\lim_{n \go \infty} \frac{\int_{0}^{t_n}A(z)dz}{t_n}-\lim_{n \go \infty}
\frac {\int_{0}^{s_n}A(z)dz}{s_n}\ \frac{s_n}{t_n} -0\\
\nonumber
& = & \rho_q-\rho_q.1 \\
& = & 0
\end{eqnarray}
Moreover,  $\lim_{n\go \infty} \frac{X(t_n)-X(s_n)}{s_n}=\lim_{n
\go \infty}\frac{X(t_n)-X(s_n)}{t_n}\ \frac{t_n}{s_n}=0$. This completes the proof.
\eProof

\begin{lemma}\rm
\label{lemleftlim}
Consider an arbitrarily fixed, increasing, unbounded time sequence
$\{t_n\}_{n=1}^{\infty}$. The following result then holds:
\eq
\nonumber
\lim_{n \go \infty}\frac{X(t_n)-X(t_n^-)}{t_n} = 0.
\en
\end{lemma}
\bProof
Clearly the result holds at times $t$ when $A(t)$ is finite. The
issue arises at times $t_n$ when $A(t_n)$ has a $\delta$-jump and the
workload suddenly shifts by a finite amount, which may actually be
increasing in consecutive jumps. Let $t_n$ be the time of a job
arrival to queue $q\in\cQ$, where $j_n$ the index of that job and
$\sigma^{j_n}_q$ the workload added by the job. It is then
sufficient to show that $\lim_{n\go \infty}
\frac{\sigma^{j_n}_q}{t_n}=0$. Indeed, note that
\begin{equation}
\label{eqalpha}
\sigma^{j_n}_q = \int_0^{t_n}A_q(t)dt - \int_0^{t_n^-}A_q(t)dt
\end{equation}
Dividing by $t_n$ and letting $n \go \infty$, we have  $\lim_{n
\go \infty} \frac{\sigma^{j_n}_q}{t_n}= \rho_q-\rho_q = 0$, which
proves the lemma. \eProof

\subsubsection{Building a Contradiction.}
\label{appfstab}

The objective of the proof is to show that $\lim_{t\go \infty}
\frac{X(t)}{t}={0}$, when $\rho\in\cR$. Since $\bB$
is a {\em positive-definite} matrix, it is sufficient to show that
$\lim_{t\go \infty}\< \frac{X(t)}{t}, \bB\frac{X(t)}{t} \> =0$.

The proof proceeds by contradiction.   Assume that $\lim \sup _{t
\go \infty} \< \frac{X(t)}{t}, \bB\frac{X(t)}{t}
\> >0$, and let $\{t_a\}_{a=1}^{\infty}$ be an increasing
unbounded time sequence on which the supremum limit is obtained;
let
\eq
\label{sdasda}
\lim_{a\go\infty}\frac{X(t_a)}{t_a} = \eta \neq 0
\en
be the corresponding limit. Such a convergent subsequence must exist
by the compactness (since bounded) of the set of possible
values\footnote{For any arrival trace, we have $\left. X(t) \leq
\int_0^t A(s)ds \right.$, which implies that $\left. \frac{X(t)}{t} \leq
\frac{\int_0^t A(s)ds}{t}
\rightarrow \rho \right.$} for $\frac{X(t)}{t}$ at large times. We will
construct a related unbounded time sequence $\{s_d\}_{d=1}^{\infty}$
and show that it has the property $\lim _{d \rightarrow \infty} \<
\frac{X(s_d)}{s_d},\bB\frac{X(s_d)}{s_d} \> >\lim_{a \rightarrow
\infty} \< \frac{X(t_a)}{t_a}, \bB\frac{X(t_a)}{t_a} \> >0$.  The
existence of such a sequence will {\em contradict} that the
supremum limit is attained on $\{t_a\}_{a=1}^{\infty}$.

We establish the required contradiction by finding an increasing
unbounded subsequence $\{t_c\}_{c=1}^{\infty}$ of
$\{t_a\}_{a=1}^{\infty}$,\ and a related sequence
$\{s_c\}_{c=1}^{\infty}$,\ which satisfy the following two {\em Key
Properties}:
\begin{enumerate}
\item[\bf I.]
$\lim_{c \rightarrow \infty}\frac{t_c-s_c}{t_c}=\epsilon \in
(0,1)$ and $s_c < t_c$ for each $c$. This implies that $\lim_{c
\go \infty} \frac{s_c}{t_c} = 1-\epsilon$.
\item[\bf II.]
$\cC(X(t))\subset \cC(\eta)$ for all $t \in (s_c, t_c]$ and each
$c$. This implies that the workload $X(t)$ drifts within the cone
$\cC(\eta)$ surrounding $\eta=\lim_{c\go\infty}
\frac{X(t_c)}{t_c}$ throughout the time interval $(s_c, t_c]$.
\end{enumerate}
The associated {\em intuition} is that $s_c$ marks the last time
before $t_c$ that the workload vector $X(s_c)$ (re)enters the cone
$\cC(\eta)$ and reaches $X(t_c)\approx\eta t_c$ at time $t_c$,
drifting in $\cC(\eta)$ throughout the time interval $(s_c,t_c]$.

Before constructing the above sequences with properties I and II
we show their implications for establishing the required
contradiction.

\begin{lemma}\rm
\label{lemlimsuptc}
If the sequences $\{t_c\}_{c=1}^{\infty}$ and
$\{s_c\}_{c=1}^{\infty}$  satisfy the {\em Properties I and II} above, then the supremum limit is not attained - as initially assumed - on the sequence $\{t_c\}_{c=1}^{\infty}$ (which is a subsequence of
$\{t_a\}_{a=1}^{\infty}$). This establishes the targeted
contradiction.
\end{lemma}

\bProof
Since $\bB$ matrix has negative or zero off diagonal elements, the  cone schedule does not apply any positive service rate to any empty queue (\ref{norate}). Therefore, by (\ref{xaxaxa}) we have
\eq
\label{eq310}
X(t_c)-X(s_c)=\int_{s_c}^{t_c}A(z)dz -\int_{s_c}^{t_c}\hat{S}(z)dz
\en
and projecting on $\bB \eta$ (\ref{sdasda}) we get
\begin{eqnarray}
\label{eqip}
\nonumber
\< X(t_c)-X(s_c),\bB {\eta} \>
& = &
\< \int_{s_c}^{t_c}A(z)dz,\bB{\eta} \> -
\< \int_{s_c}^{t_c}\hat{S}(z)dz,\bB \eta \> \\
& = &
\< \int_{s_c}^{t_c}A(z)dz,\bB{\eta} \> -
\sum_{e\in\cE} \int_{s_c}^{t_c}\< \hat{S}^e(X(z)),\bB {\eta} \>
\ind{e(z)=e} dz
\end{eqnarray}
where  $\hat{S}^e(X(z))\in\hat{\cS}^e(X(z))$ for $z\geq 0$. But because of Property II above, the workload $X(z)$ drifts in the cone $\cC(\eta)$ throughout $z\in(s_c,t_c]$, which implies that
\eq
\nonumber
\< \hat{S}^e(X(z)),\bB {\eta} \> = \max_{S\in\cS^e} \<S,\bB {\eta} \>
\en
when $e(z)=e\in\cE$. 
%{\red [Connect/refer to section om cone schedule geometry.]} 
Substituting into (\ref{eqip}) we get
\eq
\label{keyrel}
\< X(t_c)-X(s_c),\bB {\eta} \> =
\< \int_{s_c}^{t_c}A(z)dz,\bB{\eta} \> -
\sum_{e\in\cE} \left( \int_{s_c}^{t_c} \ind{e(z)=e} dz \right)
\max_{S\in\cS^e} \<S,\bB {\eta} \>
\en

Observe now that
\begin{align*}
\lim_{c\go\infty}\frac{\int_{s_c}^{t_c}A(t)dt}{t_c-s_c} 
&=
\lim_{c\go\infty}\frac{\int_{0}^{t_c}A(t)dt}{t_c}
\lim_{c\go\infty}\frac{t_c}{t_c-s_c} -
\lim_{c\go\infty}\frac{\int_{0}^{s_c}A(t)dt}{s_c}
\lim_{c\go\infty}\frac{s_c}{t_c-s_c}\\
&=
\rho\frac{1}{\epsilon} -
\rho(\frac{1}{\epsilon} - 1)\\
&=
\rho 
\end{align*}
 
because of Property I above. Dividing \ref{keyrel} by $(t_c-s_c)$ and
letting $c \go \infty$, we get
\eq
\label{dada}
\lim_{c \go \infty}\< \frac{X(t_c)-X(s_c)}{t_c-s_c}, \bB \eta \>  =
\< \rho, \bB\eta \> -
\sum_{e\in\cE} \pi^e \max_{S\in\cS^e} \<S,\bB {\eta} \> = -
\gamma(\eta) \leq 0
\en
for $\gamma(\eta)\geq 0$. The inequality
$-\gamma(\eta) = \< \rho, \bB\eta \> - \sum_{e\in\cE} \pi^e  \max_{S\in\cS^e} \<S,\bB {\eta} \> \leq 0$ is due to (\ref{lempolar}), since it is assumed that $\rho\in\cR$.

Since $\{t_c\}_{c=1}^\infty$ is a subsequence of $\{t_a\}_{a=1}^\infty$ we have $\lim_{c \go \infty} \frac{X(t_c)}{t_c} = \eta$. Using Property I and (\ref{dada}) we get the following inequality
\begin{eqnarray}
\label{eqipineq}
\nonumber
\lim_{c \go \infty} \< \frac{X(s_c)}{s_c}, \bB \eta \>
& = &
\lim_{c \go \infty}
\left\{ \< \frac{X(s_c)-X(t_c)}{s_c}, \bB \eta \> +
\< \frac{X(t_c)}{s_c}, \bB \eta \> \right\} \\
\nonumber
& = &
\lim_{c \go \infty}
\left\{\frac{t_c-s_c}{s_c}
\left[ - \< \frac{X(t_c)-X(s_c)}{t_c-s_c}, \bB \eta \> \right] +
\frac{t_c}{s_c}\< \frac{X(t_c)}{t_c}, \bB \eta \> \right\}    \\
\nonumber
& = &
\frac{\epsilon}{1 - \epsilon} \ \gamma (\eta) +
\frac{1}{1-\epsilon}\ \< \eta, \bB \eta \> \\
& >& \< \eta, \bB \eta \>
\end{eqnarray}
The last inequality is due to the facts that $\epsilon \in (0,1)$
and $ \gamma(\eta) \geq 0$.

By successive thinnings of the components of the workload vector,
we can obtain an increasing unbounded subsequence $\{s_d
\}_{d=1}^{\infty}$ of $\{s_c \}_{c=1}^{\infty}$ such that
$\lim_{d \rightarrow \infty}\frac{X(s_d)}{s_d} = \psi$ and from
(\ref{eqipineq})
\eq
\label{saasaa}
\< \psi,\bB \eta \> > \<\eta,\bB \eta \>
\en
Since $\bB$ is {\em positive-definite} we have $\< \psi- \eta,
\bB(\psi-\eta) \> \geq 0$. This implies
$\< \psi, \bB \psi \> + \< \eta, \bB \eta \> \geq \<
\psi, \bB\eta \> + \< \eta, \bB\psi \>$. Since $\bB$ is {\em
symmetric} (self-adjoint) we have $\< \eta, \bB\psi \> =
\<\psi,\bB \eta\>$. Therefore,
\eq
\nonumber
\< \psi, \bB \psi \> + \< \eta, \bB
\eta \> \geq 2 \< \psi, \bB\eta \> > 2 \<\eta,\bB \eta \>,
\en
using (\ref{saasaa}) for the last inequality.  Thus,
$\< \psi,\bB \psi \> > \<\eta,\bB \eta \>$ or
\eq
\nonumber
\lim_{d \go \infty}
\< \frac{{X}(s_d)}{s_d},\bB\frac{{X}(s_d)}{s_d}\> =
\< \psi, \bB \psi \> > \< \eta,\bB\eta \>=
{\limsup}_{t \go \infty}  \<
\frac{{X}(t)}{t},\bB\frac{{X}(t)}{t}\> > 0,
\en
giving a contradiction to the definition of ${\eta}$. This
completes the proof of Lemma \ref{lemlimsuptc}. \eProof

\subsubsection{Constructing Sequences with Properties I and II}

It now remains to construct sequences $\{t_c\}_{c=1}^\infty$ and
$\{s_c\}_{c=1}^{\infty}$ satisfying properties I and II. Their
construction is based on the intuition mentioned above, which is
made formal in the following lemma.

\begin{lemma}\rm
\label{wawa}
Suppose $\lim_{k \go \infty} \frac{X(t_k)}{t_k} = \eta
\neq 0$ for some increasing unbounded sequence
$\{t_k\}_{k=1}^{\infty}$ and nonzero ${\eta}$.  Let
\eq
\label{eqdefsc}
s_k = \sup \{t < t_k : \cC(X(t)) \nsubseteq \cC(\eta) \}
\en
be the last time before $t_k$ that the cone $\cC(X(t))$ is not
included in $\cC(\eta)$. This is the last time that $X(t)$ crosses
from outside $\cC(\eta)$ to inside, hence, $X(t)\in\cC(\eta)$ for
every $t\in(s_k,t_k]$ and the workload drifts in $\cC(\eta )$
throughout that interval. By convention $s_k=0$ if the workload
has always been in $\cC(\eta)$ before $t_k$.  We then have
\eq
\label{eqeps1}
\liminf_{k \go \infty} \frac{t_k-s_k}{t_k} =
\epsilon_1 > 0
\en
for some  $\epsilon_1 \in (0,1)$.
\end{lemma}
\bProof
Arguing by contradiction, suppose that there exists an increasing
unbounded subsequence $\{t_n\}_{n=1}^\infty$ of
$\{t_k\}_{k=1}^\infty$ such that $\lim_{n \go \infty}
\frac{t_n-s_n}{t_n}=0$. From Lemma \ref{lemwkld} we have that
$\lim_{n\go \infty} \frac{X(t_n)-X(s_n)}{s_n}=0$. Since
$\lim_{n\go \infty} \frac{X(t_n)}{t_n}=\eta$, we then get $\lim_{n
\go\infty}\frac{X(s_n)}{s_n} = \eta$. Further, to allow for the
possibility of job arrival that instantaneously shifts the workload
from outside $\cC(\eta)$ to inside, we note from Lemma
\ref{lemleftlim} that we have $\lim_{n \go \infty}\frac{X(s_n^-)}{s_n} =
\eta$. But according to the definition of $s_n$ the workload
$X(s_n)$ must be outside $\cC(\eta)$, so $\lim_{n
\go \infty}\frac{X(s_n^-)}{s_n}$ could not converge to $\eta$.
This establishes the necessary contradiction, showing
\ref{eqeps1} and completing the proof of the Lemma \ref{wawa}. \eProof

We are now ready to construct sequence $\{s_c\}_{c=1}^{\infty}$
satisfying properties I and II. We rename the sequence defined in
(\ref{eqdefsc}) to be $\{\hat{s}_c\}$ and choose
$s_c=\max\{\hat{s}_c, (1-\epsilon_2)t_c \}$, for some
$\epsilon_2\in(0,1)$. (The second term $(1-\epsilon_2)t_c$ is used
to guard against the degenerate case where $\hat{s}_c$ is finite
because the workload $X(t)$ is always in $\cC(\eta )$ after some
finite time.) Then we have the properties:

\begin{enumerate}
\item
$\lim_{c \rightarrow \infty}\frac{t_c-s_c}{t_c}=\epsilon \in
(0,1)$ and $s_c < t_c$ for each (large) $c$.
\item
$C(X(t))\subset C(\eta)$ for all $t \in (s_c, t_c]$ and each
(large) $c$.
\end{enumerate}
This means that $\{s_c\}_{c=1}^{\infty}$ and
$\{t_c\}_{c=1}^{\infty}$  satisfy both Properties I and II, and
Lemma \ref{lemlimsuptc} completes the proof of rate stability in
Theorem \ref{thrate}. \eProof

\section{Performance Issues}
\label{secPerf}

Section \ref{secStability} established the universal stability of cone schedules for an entire class of matrices $\bB$. A natural question to consider is how the selection of $\bB$  from within this class of matrices will affect other performance measures such as average workload and waiting time.

In \cite{Stol:04} it was shown that  for a similar queueing system (with one environment state and nonnegative $S_q$ vectors), the class of {\em MaxWeight} schedules, which are equivalent to cone schedules with a diagonal $\bB$ matrix, will minimize the total workload in the system as well as the holding cost rate asymptotically in heavy traffic.
While a formal proof of a corresponding result is beyond the scope of this paper, we conjecture that a similar result will hold for these generalized cone schedules. This can be observed by considering the limiting behavior of the cone schedules when $X$ is large.
\begin{multline}
\< X(t^+), \bB, X(t^+) \> = \\  
 \< X(t), \bB, X(t) \> +  \< A(t) - S(t), \bB (A(t)-S(t)) \> +  2 \< A(t), \bB X(t) \>  -  2 \< S(t), \bB X(t)\>
\end{multline}

for $t^+>t$ the workload immediately after time $t$. 
If $X(t)$ is large and fixed, then minimizing the expectation of the above equation is equivalent to maximizing $ \< X(t), \bB S(t)\>$, because the first term is fixed by $X(t)$ and no other terms grow with $X(t)$. This causes us to conjecture that the schedule that minimizes $\lim_{t\go\infty}\frac{1}{t} \int_0^t\<X(z) , \bB X(z) \>$ will be the cone schedule with matrix $\bB$.  A full proof of this optimality would require considerably more restriction on the arrival trace than has been presented in this paper.

\begin{figure}[htbp]
\begin{center}

\includegraphics[width=0.3\linewidth]{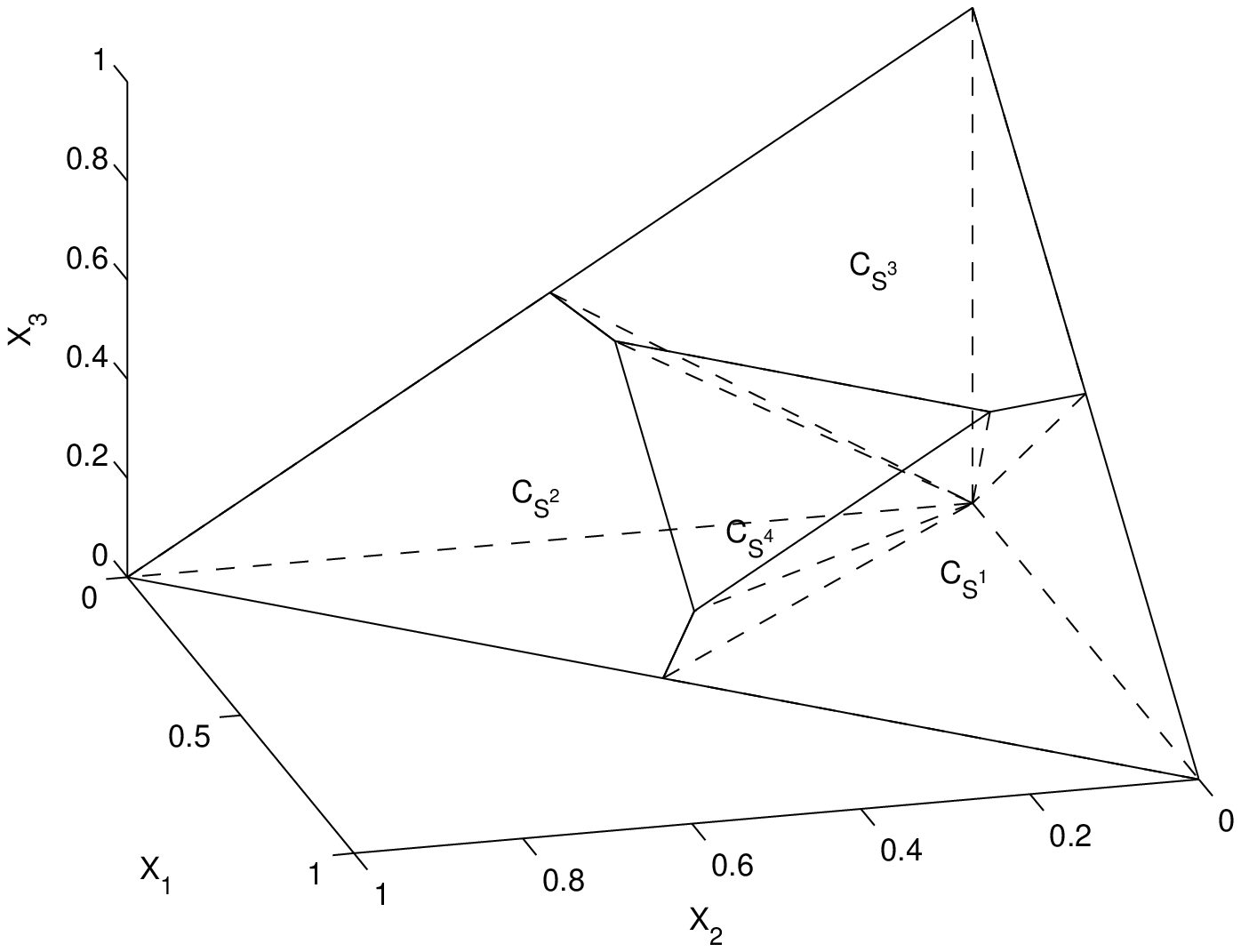}
\includegraphics[width=0.3\linewidth]{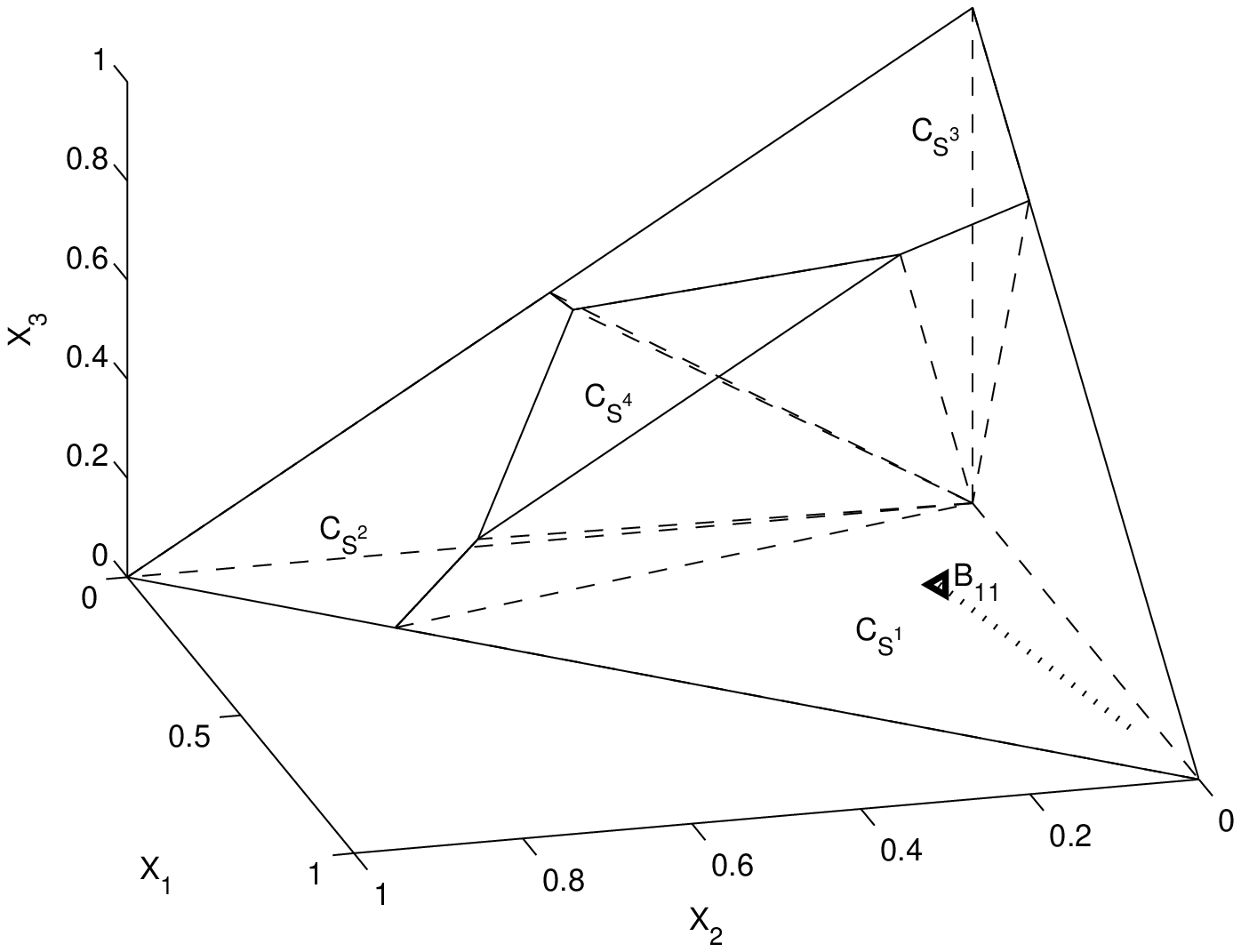}
\includegraphics[width=0.3\linewidth]{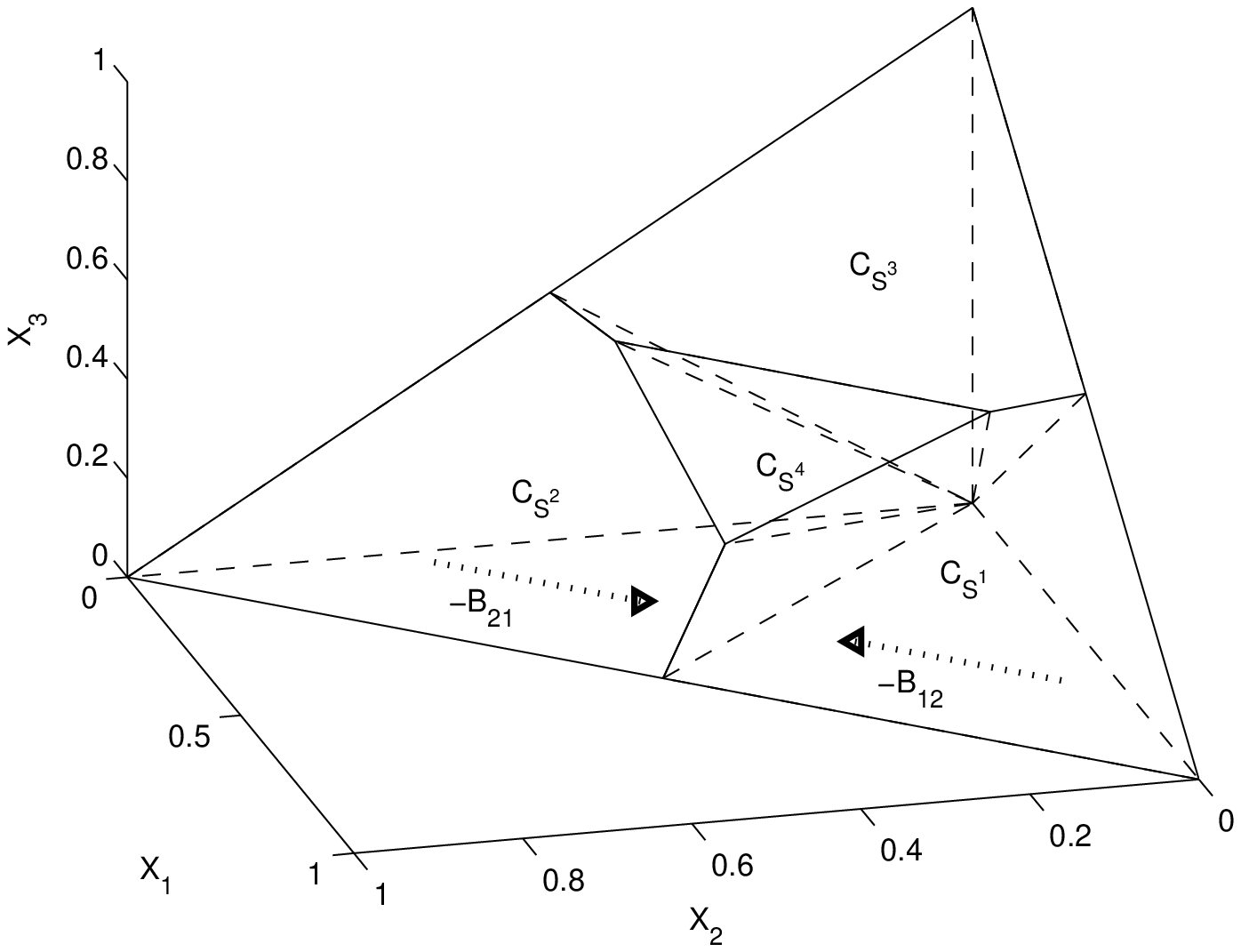}
\caption{{\bf The cone space with varying $\bB$ matrices}. The plots above illustrate the impact of matrix $\bB$ on the cone space for a  system with $Q=3$ queues and $M=4$ service vectors. The first plot shows the cones for an identity matrix $\bB$. The middle plot has a diagonal $\bB$ with large positive weight on the first queue, focusing more attention on service vectors which serve that queue. The third plot shows the impact of off diagonal elements of $\bB$. These elements affect the boundary between two queues. }
\label{figVarB}
\end{center}
\end{figure}

From a geometric point of view, the cones $\cC_m^e$ shift (and
expand or contract), as the weights assigned to particular queues
are adjusted. Figure \ref{figVarB} illustrates a simple system where the matrix $\bB$ transforms the cone space. The diagonal elements of $\bB$ expand and contract the cones in the dimension of the corresponding queue. The off-diagonal elements move the boundary between adjacent cones where both cones have a nonzero service rate to the two corresponding queues.

As highlighted in the stability proof, as $X$ increases in magnitude in any given direction, the service vectors applied will rotate through the vectors which maximize $\< S, \bB X \>$ for each environment. While in a particular environment, $X$ will  be drawn toward the boundary of its current cone. In particular, if no new arrivals were allowed and the environment stayed constant, then $X$ would follow a deterministic path to a cone boundary. Once at a boundary, the cone schedule would fluctuate around the service vectors which are optimal at that boundary and the workload would be drawn down along that boundary. The boundary planes act as attractors for the balance of queues in the system.
Therefore one can view the matrix $\bB$ as transforming the cone space in order to set the appropriate attracting boundary planes given by the cone intersections. This leads to an understanding of $\bB$ as an important control on the relative importance of different workload dimensions.

Cone schedules perform constrained
{\em dynamic load balancing} of the queue workloads (weighted by the
elements of $\bB$), observing the service constraints encoded in the
service vectors $\cS$. As the workload of a queue increases
excessively, the schedule shifts attention to it and selects available
service configurations $S\in\cS^e$ that provide more service capacity to that queue,
potentially at
the expense of others. That lowers the workload at the queue, trading
it for increased workload in others and load balancing them.

A strictly diagonal matrix $\bB$ induces a direct {\em simple
priority scheme}. That is, as the weight $B_{qq}$ of queue $q$ is
increased (while those of others remain constant), the queue
attains higher service priority. This results in the queue
receiving more service bandwidth over time and enjoying a lower
workload.

When $\bB$ has negative off-diagonal elements, those have an
indirect effect on service priorities, entangling the queues and
inducing a {\em coupled priority scheme}. That is, when $B_{pq}<0$
with $p\neq q$,
the relative priority of queue $p$ \ decreases as the workload of
queue $q$ increases.
As $X_q$ grows in size, more attention needs to be paid in servicing queue $q$,
while as $X_p$ grows in size, less attention is paid to queue $q$. It can be seen
that the weight $B_{pq}<0$ induces a specific coupling between the corresponding
queues.

We also observe that the proof of stability in section \ref{secProof} is robust to any sublinear perturbations of information or time. In particular, if there is a switching delay between configurations or an information lag in knowledge of $X(t)$, or some error in the calculation of $\max_S \<S, \bB X \>$ then {\em as long as the corresponding perturbation does not grow linearly with $t$, then rate stability will still be assured}. See \cite{RoB:09} for more detail on this observation. For example, if any calculation error or delay is bounded, then stability will hold. For clarity in the proofs we have not included additional terms, but the intuition is that any such sublinear term  will have no impact on the limiting case as $X(t)$ becomes large. 

For some processing systems, there can be computational issues in the requirement to calculate $\max_S \<S, \bB X\>$ in real-time over every possible service vector. The geometric structure of the cone schedules helps to overcome this by recognizing that {\em when the workload vector $X$ is large, any bounded change in workload will move the workload between adjacent cones} (adjacent cones have a common boundary when $\Re^Q$ is divided into the cones $\cC_S$). Therefore, cone schedules can be implemented by  evaluating $\<S, \bB X \>$ over a much smaller subset of service vectors at each point in time. This was shown to be a special case of sublinear perturbations in the cone schedules, and discussed in detail in \cite{RoB:09}.

To conclude, the adjustment of the $Q \times Q$ entries of the
matrix $\bB$ allows for generating a rich family of stable service
schedules. The dynamic priorities of the queues relate directly to
the quality of service (QoS) they receive and the workload they
see. We are currently exploring such performance issues further.

\section{Conclusions and Further Research}
\label{secConclusions}

We have established that the family of Cone  Schedules maximizes the
system throughput for very general processing systems under very general conditions. 
These schedules naturally select the best
available vectors, and this leads to the maximum possible system
throughput.  Arrival and service processes are as general as
possible, and the stability proofs are presented with minimal assumptions.  
 Previous stability results are generalized here to a setting with generalized service vectors, fluctuating resource availability and continuous time scheduling.

By exploring the analysis from a geometric standpoint we have
gleaned important intuition for stability as well as performance and scalability of the schedules. Further research is necessary to deeper explore the
effects of changes to the $\bB$ matrix.  The transition between
environments is also a topic for further study, since it may not be
trivial, especially as switching costs become important.

\bibliographystyle{apalike}

\bibliography{masterbib}

% \received{March 2011}{March 2011}

%\end{spacing}

\end{document}